\documentstyle[a4p,12pt,epsfig]{article}

\begin{document}
\newcommand {\epem}     {e$^+$e$^-$}
\newcommand {\durham}   {$k_\perp$}
\newcommand {\rcone}    {$R$}
\newcommand {\econe}    {$\epsilon$}
\newcommand {\qq} {q$\overline{\mathrm{q}}$}
\newcommand {\gluglu} {gg}
\newcommand {\gincl} {g$_{\,\mathrm{incl.}}$}
\newcommand {\xmax} {$x_{\mathrm{max}}$}
\newcommand {\mngincl} {$\langle n_{\,\mathrm{ch.}} 
      \rangle_{\mathrm{g}_{\,\mathrm{incl.}}}$}
\newcommand {\mnuds} 
   {$\langle n_{\,\mathrm{ch.}} \rangle_{\mathrm{uds\,hemis.}}$}
\newcommand {\mnudsmod}
   {$\langle n_{\,\mathrm{ch.}} 
   \rangle_{\mathrm{uds\,hemis.}}^{41.8\,\mathrm{GeV}}$}
\newcommand {\egincl} {$\langle E 
      \rangle_{\mathrm{g}_{\,\mathrm{incl.}}}$}
\newcommand {\ecm} {\mbox{$E_{\mathrm{c.m.}}$}}
\newcommand {\lms} {$\Lambda_{\overline{\mathrm{MS}}}$}
\newcommand {\ejet} {$E_{\,\mathrm{jet}}$}
\newcommand {\rch} {$r_{\,\mathrm{ch.}}$}
\newcommand {\nch} {$n_{\,\mathrm{ch.}}$}
\newcommand {\mnch} {$\langle n_{\,\mathrm{ch.}} \rangle$}
\newcommand {\mnchmath} {\langle n_{\,\mathrm{ch.}} \rangle}
\newcommand {\fq} {F_q}
\newcommand {\kq} {K_q}
\newcommand {\hq} {H_q}
\newcommand {\rfq} {r_{F_q}}
\newcommand {\rkq} {r_{K_q}}
\newcommand {\rhq} {r_{H_q}}
\newcommand {\hqgincl} { (H_q)_{\mathrm{g}_{\,\mathrm{incl.}}} }
\newcommand {\hqgluon} { (H_q)_{\mathrm{gluon}} }
\newcommand {\hquds} { (H_q)_{\mathrm{uds\,hemis.}} }
\newcommand {\hqudsmod} { (H_q)_{\mathrm{uds\,hemis.}}^{41.8\,\mathrm{GeV}} }
\newcommand {\lmsnffive} {$\Lambda_{\overline{\mathrm{MS}}}^{(n_f=5)}$}
\newcommand {\lmsnfthree} {$\Lambda_{\overline{\mathrm{MS}}}^{(n_f=3)}$}
\newcommand {\pminch} {$p_{\mathrm{min.}}^{\,\mathrm{ch.}}$}
\newcommand {\qzero} {$Q_0$}
\begin{titlepage}
\noindent
\begin{center}
  {\large   EUROPEAN LABORATORY FOR PARTICLE PHYSICS }
\end{center}

\bigskip
\begin{tabbing}
\` CERN-PPE/97-105 \\
\` 1 August 1997 \\
\end{tabbing}

\bigskip\bigskip
\bigskip\bigskip
 
\begin{center}{\Large\bf
Multiplicity distributions of
gluon and quark jets and tests of
QCD analytic predictions
}
\end{center}

\begin{center}
{\large
The OPAL Collaboration
}
\end{center}
\begin{center}{\large\bf  Abstract}\end{center}
\bigskip
\noindent
Gluon jets are identified in {\epem} hadronic 
annihilation events by tagging two quark jets in 
the same hemisphere of an event.
The gluon jet is defined inclusively
as all the particles in the opposite hemisphere.
Gluon jets defined in this manner
have a close correspondence to gluon jets as
they are defined for analytic calculations,
and are almost independent of a
jet finding algorithm.
The charged particle multiplicity distribution
of the gluon jets is presented,
and is analyzed for its mean, dispersion, skew,
and curtosis values, 
and for its factorial and cumulant moments.
The results are compared to the analogous
results found for a sample of light quark (uds) jets,
also defined inclusively.
We observe differences between the mean,
skew and curtosis values of gluon and quark jets,
but not between their dispersions.
The cumulant moment results are compared to the
predictions of QCD analytic calculations.
A calculation which includes
next-to-next-to-leading order corrections
and energy conservation
is observed to provide a much improved description
of the data compared to a next-to-leading order
calculation without energy conservation.
There is agreement between the data and
calculations for the ratios of the cumulant moments
between gluon and quark jets.

\bigskip\bigskip\bigskip
\begin{center}{\large
(To be submitted to Zeitschrift f\"{u}r Physik C)
}\end{center}

\end{titlepage}
\clearpage

{\small 
\begin{center}{{\Large   The OPAL Collaboration} \\
\vspace*{0.5cm}
K.\thinspace Ackerstaff$^{  8}$,
G.\thinspace Alexander$^{ 23}$,
J.\thinspace Allison$^{ 16}$,
N.\thinspace Altekamp$^{  5}$,
K.J.\thinspace Anderson$^{  9}$,
S.\thinspace Anderson$^{ 12}$,
S.\thinspace Arcelli$^{  2}$,
S.\thinspace Asai$^{ 24}$,
D.\thinspace Axen$^{ 29}$,
G.\thinspace Azuelos$^{ 18,  a}$
% ...
}\end{center}
\vspace*{-.4cm}
% ...
$^{  2}$Dipartimento di Fisica dell' Universit\`a di Bologna and INFN,
I-40126 Bologna, Italy
\newline
% ...
$^{  5}$Cavendish Laboratory, Cambridge CB3 0HE, UK
\newline
% ...
$^{  8}$CERN, European Organisation for Particle Physics,
CH-1211 Geneva 23, Switzerland
\newline
$^{  9}$Enrico Fermi Institute and Department of Physics,
University of Chicago, Chicago IL 60637, USA
\newline
% ...
$^{ 12}$Indiana University, Department of Physics,
Swain Hall West 117, Bloomington IN 47405, USA
\newline
% ...
$^{ 16}$Department of Physics, Schuster Laboratory, The University,
Manchester M13 9PL, UK
\newline
% ...
$^{ 18}$Laboratoire de Physique Nucl\'eaire, Universit\'e de Montr\'eal,
Montr\'eal, Quebec H3C 3J7, Canada
\newline
% ...
$^{ 23}$Department of Physics and Astronomy, Tel Aviv University,
Tel Aviv 69978, Israel
\newline
$^{ 24}$International Centre for Elementary Particle Physics and
Department of Physics, University of Tokyo, Tokyo 113, and
Kobe University, Kobe 657, Japan
\newline
% ...
$^{ 29}$University of British Columbia, Department of Physics,
Vancouver BC V6T 1Z1, Canada
\newline
% ...
$^{  a}$ and at TRIUMF, Vancouver, Canada V6T 2A3
\newline
% ...
$^{  c}$ and Institute of Nuclear Research, Debrecen, Hungary
\newline
% ...
}

\clearpage\newpage
\section{Introduction}
\label{sec-intro}

Many experimental studies of quark jets
have been performed at {\epem} colliders.
Such studies are natural,
since hadronic events in {\epem} annihilations
above the $\Upsilon$ region and below the
threshold for W$^+$W$^-$ production are
believed to arise uniquely from the point-like 
creation of quark-antiquark {\qq} pairs.
Production of the {\qq} pair
from a color-singlet point source
allows the quark jets to be defined inclusively,
by sums over the particles in
an event or the event hemispheres.
In contrast,
conclusive experimental studies of gluon jets 
have been rare.
This is because the
creation of a gluon jet pair, {\gluglu},
from a color singlet point source
--~allowing an inclusive definition
analogous to that described above for quark jets~--
has been only rarely observed in 
nature.\footnote{It is possible
to identify a pure source of {\gluglu} events in
radiative $\Upsilon$ decays,
such as in $\Upsilon$(3S)$\,\rightarrow
\gamma\chi_{\mathrm{b}}^{\prime}$
followed by $\chi_{\mathrm{b}}^{\prime}
\rightarrow\,${\gluglu}~\cite{bib-cleoqg};
however,
the jet energies are only about 5~GeV in this case,
which limits their usefulness for jet studies.}
In most studies of gluon jets at {\epem} colliders,
a jet finding algorithm is used to
select an exclusive sample of three-jet {\qq}g events.
The same jet finder is used
to artificially divide the particles
of an event into a gluon jet part and two quark jet parts.
In general,
the results depend strongly on the algorithm chosen.
Furthermore, use of a jet finder precludes
a quantitative test of QCD analytic predictions
for gluon and quark jet properties.
For the analytic calculations,
the gluon and quark jet characteristics are
given by inclusive sums over the particles in
color singlet {\gluglu} and {\qq} events, respectively,
as described above.
Thus, the theoretical results are {\it not} 
restricted to three-jet events defined by a jet finder
and do {\it not} employ a jet finder 
to assign particles to the jets.

In~\cite{bib-jwg},
a method was proposed for LEP experiments
to identify gluon jets using an inclusive
definition similar to that used for analytic calculations.
The method is based on rare events
of the type {\epem}$\rightarrow\,${\qq}$\,${\gincl},
in which the q and $\overline{\mathrm{q}}$ are
identified quark (or antiquark) jets which appear
in the same hemisphere of an event.
The object {\gincl}, taken to be the gluon jet,
is defined by the sum of all particles observed in the
hemisphere opposite to that containing 
the q and~$\overline{\mathrm{q}}$.
In the limit that the q and~$\overline{\mathrm{q}}$ 
are collinear,
the gluon jet {\gincl} is produced under the same
conditions as gluon jets in {\gluglu} events.
The {\gincl} jets therefore correspond closely to 
single gluon jets in {\gluglu} events,
defined by dividing the {\gluglu} events in half
using the plane perpendicular to the principal event axis.
First experimental results using this method were
presented in~\cite{bib-opalqg96}.

The results in~\cite{bib-opalqg96} were limited
to the mean charged particle multiplicity values
of gluon and quark jets.
In this paper,
we extend this study to include the
full multiplicity distributions.
The data were collected using the OPAL detector at LEP.
For the quark jet sample,
we select light quark (uds) event hemispheres,
as in~\cite{bib-opalqg96}.
Use of light quark events results in
a better correspondence between the data
and the massless quark assumption
employed for analytic calculations,
while use of event hemispheres to define the quark jets
yields an inclusive definition analogous to that
of the gluon jets {\gincl}.
The multiplicity distributions of the gluon and quark jets
are analyzed for their mean, dispersion, skew
and curtosis values.
In addition,
we perform a factorial moment analysis of the 
gluon and quark jet multiplicity distributions
in order to test the predictions of QCD analytic 
calculations~\cite{bib-webber,bib-analytic1}
of those moments.

\section{Detector and data sample}
\label{sec-detector}

The OPAL detector is described in
detail elsewhere~\cite{bib-detector}.
The present analysis is based on a sample of
about $3\,708\,000$ hadronic Z$^0$ decay events
collected by OPAL from 1991 to 1995.
Charged tracks measured in the OPAL central detector and
clusters of energy measured in the electromagnetic calorimeter
were selected for the analysis using the criteria given
in~\cite{bib-qg95b}.
To minimize double counting of energy,
clusters were used only if they were
not associated with a charged track.
Each accepted track and unassociated cluster
was considered to be a particle.
Tracks were assigned the pion mass.
Clusters were assigned zero mass since they originate
mostly from photons.
To eliminate residual background and events
in which a significant number of particles was lost
near the beam direction,
the number of accepted charged tracks was required
to be at least five and the thrust axis~\cite{bib-thrust}
of the event,
calculated using the particles,
was required to satisfy
$|\cos (\theta_{\mathrm{thrust}})|<0.9$,
where $\theta_{\mathrm{thrust}}$ is the
angle between the thrust and beam axes.
The residual background from all sources
was estimated to be less than~1\%.

\section{Gluon jet selection}
\label{sec-gluon}

For this study,
a gluon jet is defined inclusively
by the particles observed in an {\epem} event hemisphere
opposite to a hemisphere containing
an identified quark and antiquark jet,
as stated in the introduction.
The selection of inclusive gluon jets, {\gincl},
is performed using the technique
presented in~\cite{bib-opalqg96}.
More details concerning the motivation for
the selection choices are given there.
To select the {\gincl} gluon jets,
each event is divided into hemispheres using the
plane perpendicular to the thrust axis.
Exactly two jets are reconstructed in each hemisphere,
using the {\durham} (``Durham'')
jet finder~\cite{bib-durham}.
The results for the gluon jet properties are
almost entirely insensitive to this choice
of jet finder,
as is discussed in~\cite{bib-jwg}
and below in sections~\ref{sec-ggcompare}
and~\ref{sec-systematic}.
Next, we attempt to reconstruct a displaced
secondary vertex in each of the four jets.
Displaced secondary vertices are associated with 
heavy quark decay,
especially that of the b quark.
At LEP, b quarks are produced almost exclusively
at the electroweak vertex:
thus a jet containing a b hadron is almost always a quark jet.
To identify secondary vertices in jets,
we employ the method given in~\cite{bib-qg95a}.
Briefly, a secondary vertex is required to contain at least
three tracks,
at least two of which have a signed impact
parameter value in the $r$-$\phi$ plane\footnote{Our
coordinate system is defined so that
$z$~is the coordinate parallel to the e$^-$ beam axis,
$r$~is the coordinate normal to the beam axis,
$\phi$~is the azimuthal angle around the beam axis and
$\theta$~is the polar angle \mbox{with respect to~$z$.}}
with respect to the primary event vertex,
$b$,
which satisfies $b/\sigma_b>2.5$,
with $\sigma_b$ the error of $b$.
For jets with such a secondary vertex,
the signed decay length, $L$,
is calculated with respect to the primary vertex,
along with its error,~$\sigma_L$.
To be tagged as a quark jet,
a jet is required to have a visible energy of at least 5~GeV
and a successfully reconstructed secondary vertex
with $L<2.0$~cm and
$L/\sigma_L>3.5$.\footnote{In~\cite{bib-opalqg96},
we utilized a more stringent requirement of $L/\sigma_L>5.0$.}
The visible energy of a jet is defined by the sum of
the energy of the particles assigned to the jet.
We refer to a hemisphere with two tagged jets
as a tagged hemisphere.

We next examine the angles that the two jets
in a tagged hemisphere
make with respect to the thrust axis and to each other.
If the two jets are close together,
or if one of the two jets is much more energetic
than the other,
it is very likely that one of the two jets is
a gluon jet due to the strong kinematic similarity
to an event with gluon radiation from a quark
and because of the finite probability
for a gluon jet to be identified
as a b quark jet (see below).
To reduce this background,
we require the angle between the jets and thrust axis
to exceed~10$^\circ$ and the angle
between the two jets to exceed~50$^\circ$.
A last requirement is that the two jets lie no
more than 70$^\circ$ from the thrust axis in order
to eliminate jets near the hemisphere boundary.
In total,
324 events are selected for the final gluon jet
{\gincl} sample.\footnote{Because we use a somewhat 
different event selection here,
and because our data have been reprocessed using improved
detector calibrations since the time of
our previous publication~\cite{bib-opalqg96},
only 60\% of these events are in common with
the {\gincl} sample in~\cite{bib-opalqg96}.}
There are no events in which both hemispheres
are tagged.

We estimate the purity of this sample using the
Jetset parton shower Monte Carlo~\cite{bib-jetset},
including detector simulation~\cite{bib-gopal} 
and the same analysis
procedures as are applied to the data.
The Jetset sample is a combination of events generated
using version 7.3 of the program with the
parameter values given in~\cite{bib-qg93}
and of events generated using version~7.4 of 
the program with the
parameter values given in~\cite{bib-qg95b}.
The initial Monte Carlo samples have about $3\,000\,000$ events
for version 7.3 and $4\,000\,000$ events for version~7.4.
The two Jetset versions yield results which are consistent
with each other to within the statistical uncertainties
and so we combine them.
Using the Jetset events,
the hadron level jets are examined
to determine whether they are associated with an
underlying quark or antiquark jet.
To perform this association,
the Monte Carlo events are also examined at the parton level.
We determine the directions of the primary
quark or antiquark from the Z$^0$ 
decay after the parton shower evolution
has terminated.
The hadron jet closest to 
the direction of an evolved primary quark or antiquark
is considered to be a quark jet.
The distinct hadron jet closest to the evolved primary
quark or antiquark not associated with this first
hadron jet is considered to be the other quark jet.
With the final cuts,
Jetset predicts that both jets in the tagged hemisphere
are quark jets with $(80.5\pm 1.6)$\% probability,
where the uncertainty is statistical:
this is the estimated purity of the {\gincl} gluon jet sample.
As an alternative method to estimate the gluon jet purity,
we determine the fraction of events in the Monte Carlo 
{\gincl} sample for which the 
evolved primary quark and antiquark
are both in the hemisphere opposite the {\gincl} jet:
this yields the same estimate as given above.
The Monte Carlo predicts that about 80\% of the background
is comprised of b events in which a gluon jet
is mistakenly tagged as a b jet,
while the other 20\% is comprised about evenly
of u, d, s and c events in which both a
quark jet and a gluon jet are mistakenly
tagged as b jets.
The background events occur mostly when two tracks
from the decay of a $\Lambda$ or K$_{\mathrm S}^0$
hadron are combined
with a third track to define a secondary vertex.

Because we rely on displaced secondary vertices
to identify quark jets,
the {\gincl} jets in our study are 
contained in heavy quark events.
The Monte Carlo with detector simulation predicts that
about 95\% of the events in the 
{\gincl} sample are b events.
This reliance on b events is not expected to affect our results
since the properties of hard, acollinear gluon jets
do not depend on the event flavor according to QCD.
More details are given in~\cite{bib-opalqg96}.

We note that the {\gincl} jet tag rate,
defined by the ratio of the number of {\gincl} jets
to the number of events in the initial
inclusive multihadronic event sample, is
$(8.74\pm 0.49\,\mathrm{(stat.)})\times 10^{-5}$
for the data and 
$(8.83\pm 0.36\,\mathrm{(stat.)})\times 10^{-5}$
for the Monte Carlo.
Thus the Monte Carlo reproduces the measured tag
rate well.

The mean energy of the gluon jets, {\egincl},
is less than the beam energy because the two quark jets
against which {\gincl} recoils are not entirely collinear.
The mean visible energy of the {\gincl} jets,
corrected for the effects of the detector and
initial-state photon radiation,
is $41.8\pm 0.6\,\mathrm{(stat.)}$~GeV.
As an alternative method to estimate the gluon jet energy,
we employ the technique of calculated jet energies
for massive jets.
A jet direction is determined for the gluon jet by
summing the momenta of the particles in the
{\gincl} hemisphere.
The angles between the {\gincl} jet and the
two jets in the tagged hemisphere are used in
conjunction with the measured jet velocities
to calculate the jet energies assuming 
energy-momentum conservation.\footnote{This method 
results in a better estimate of the
{\gincl} jet energy than the method assuming
massless jets which we employed 
in~\cite{bib-opalqg96}:
the Monte Carlo without detector simulation
yields identical results
for the visible and calculated {\gincl} jet
energies if the massive formula is used,
whereas the calculated energy is about 2~GeV
smaller than the visible one
if the massless formula is used.}
The velocity of a jet is given by the magnitude
of its visible 3-momentum divided by its visible energy.
Using this method,
the mean gluon jet energy is determined to be
$41.5\pm 0.3\,\mathrm{(stat.)}$~GeV,
which is consistent with the result given
above for the visible energy.
In this paper,
we choose to use the visible {\gincl} energy,
rather than the calculated energy,
because it corresponds more closely 
to the jet energy as it is defined for the uds hemisphere
quark jets.
The difference between the mean visible
and calculated jet energies is used to define a
systematic uncertainty.
The mean energy of the gluon jets in our study is
therefore
{\egincl}=$41.8\pm 0.6\,\mathrm{(stat.)}
\pm 0.3\,\mathrm{(syst.)}$~GeV.

\section{Monte Carlo comparison of {\boldmath{\gincl}} and
{\boldmath{\gluglu}} jets}
\label{sec-ggcompare}

Our analysis of gluon jets is based on the premise
that {\gincl} jets from {\epem}
annihilations are equivalent to
hemispheres of {\gluglu} events produced from
a color singlet point source,
with the hemispheres defined
by the plane perpendicular to the thrust axis.
Although high energy {\gluglu} events are not
available experimentally,
they may be generated using a 
QCD Monte Carlo event generator.
The viability of our premise can be tested
by comparing the Monte Carlo predictions for {\gluglu} 
event hemispheres and {\gincl} jets.
Such a comparison has already been
presented in~\cite{bib-jwg}
for the mean charged particle multiplicity values,~{\mnch}
(see also~\cite{bib-jwgfaro}).
Here, we extend this comparison to the
full multiplicity distribution,
P({\nch}) versus~{\nch},
with P({\nch}) the probability that an event will
be observed with a charged particle multiplicity~{\nch}.

\begin{figure}[ht]
\epsfxsize=15cm
\epsffile[15   100 545 700]{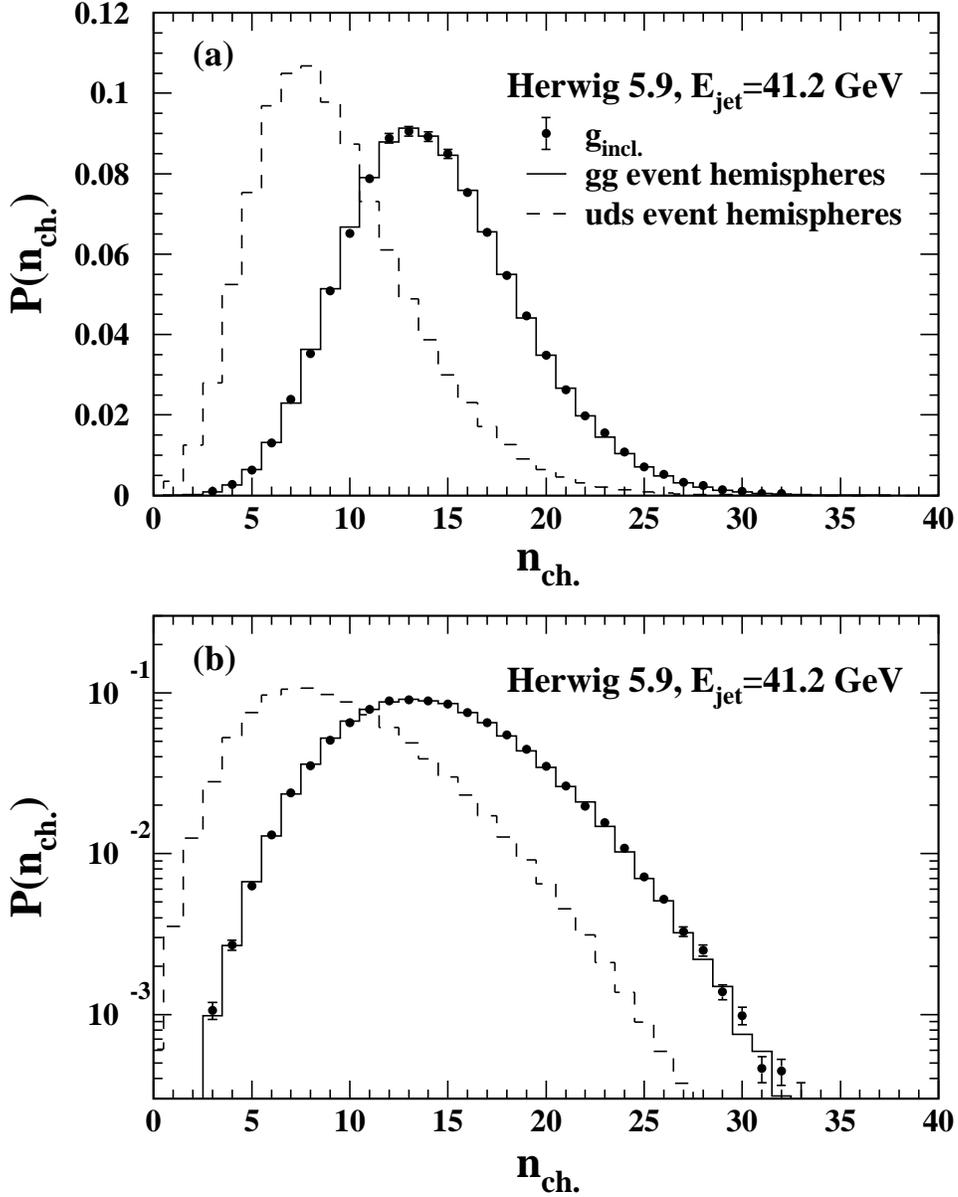}
\caption{
The prediction of the Herwig parton shower
event generator for the charged particle
multiplicity distribution of {\gincl} gluon jets 
from {\epem} annihilations,
in comparison to the Herwig prediction for
{\gluglu} and uds event hemispheres:
(a)~on a linear scale, and (b)~on a logarithmic scale.
The jet energies are 41.2~GeV,
corresponding to a c.m. energy
of 91.2~GeV for the generation of the
{\epem}$\rightarrow\,${\qq}$\,${\gincl} events.
}
   \label{fig-mcnch91}
\end{figure}

The points with error bars in Figure~\ref{fig-mcnch91}
show the prediction of the Herwig parton shower
Monte Carlo~\cite{bib-herwig}, version~5.9,
for the charged particle multiplicity distribution
of {\gincl} jets.
The uncertainties are statistical.
The parameter set we use is the same as that
given in~\cite{bib-qg95b} for Herwig, version~5.8,
except that the value of the cluster
mass cutoff CLMAX has been increased from 3.40~GeV/$c^2$
to 3.75~GeV/$c^2$
to improve the model's description of~{\mnch}
in inclusive hadronic Z$^0$ decays.
The {\epem}$\rightarrow\,${\qq}$\,${\gincl} events
were generated using a center-of-mass (c.m.) energy, 
{\ecm}, of 91.2~GeV to correspond to the data.
The {\gincl} identification
was performed using the same procedure as is
described for the data in section~\ref{sec-gluon},
except that the two quark jets against which
the {\gincl} jet recoils were identified
using the Monte Carlo method described in 
section~\ref{sec-gluon}.
In particular,
the angular cuts on the directions of the quark jets with
respect to the thrust axis and to each other have been applied.
The resulting mean energy of the Monte Carlo
{\gincl} jets is~41.2~GeV
with a negligible statistical uncertainty.

Shown by the solid histogram in Figure~\ref{fig-mcnch91}
is the prediction of Herwig for {\gluglu} event hemispheres.
The {\gluglu} events were generated using a c.m. energy
of 82.4~GeV so that
the hemisphere energies are the same as for the {\gincl} jets.
It is seen that the results for the {\gincl} jets 
and the {\gluglu} event hemispheres are essentially identical.
This establishes the viability of our method,
confirming the results of~\cite{bib-jwg}.
Similar agreement between the predicted multiplicity
distributions of {\gincl} jets and {\gluglu} event hemispheres 
is obtained if Jetset is used to generate the
samples rather than Herwig,
or if the JADE-E0~\cite{bib-jade}
or cone~\cite{bib-opalcone}
jet finder is used to identify the quark jets
for the {\gincl} jet selection,
rather than the {\durham} jet finder.

\section {uds quark jet selection}
\label{sec-uds}

The uds quark jets in our study are defined inclusively,
by summing the particles observed in an event hemisphere
opposite to a hemisphere containing an identified uds jet.
Since there are only 324 gluon jets in our study,
it is not necessary to use the entire data sample
of about $3\,708\,000$ events
mentioned in section~\ref{sec-detector} 
for the uds jet analysis.
Instead,
we base the uds jet selection on an initial sample of
about $396\,000$ hadronic annihilation
events with c.m. energies within~100~MeV of the Z$^0$ peak.

To select the uds jets,
we divide each event into hemispheres
using the plane perpendicular to the thrust axis.
Selection criteria are applied to each hemisphere
separately using charged tracks that appear in a cone 
of half angle 40$^\circ$ around the thrust axis.
The reason for the restriction
to tracks which lie within 40$^\circ$ of the thrust axis
is to avoid using tracks near the hemisphere boundary.
An algorithm is applied to identify charged tracks which are
consistent with arising from photon 
conversions~\cite{bib-idgcon}.
Removing such tracks from consideration,
the number of tracks in the cone
which have a signed impact parameter significance,
$b/\sigma_b$,
greater than 1.5 is determined.
A hemisphere is tagged as containing a uds jet
if the number of tracks in the cone which have
$b/\sigma_b>1.5$ is zero.
In total, $188\,288$ hemispheres are tagged.
This number includes $30\,303$
events for which both hemispheres are tagged.
The estimated uds purity of this sample,
obtained by treating Jetset events with detector simulation
in the same manner as the data, is~80.9\%,
with a negligible statistical uncertainty.
The Monte Carlo predicts that about
70\% of the background
events are c events and that 30\% are b events.
The uds jet tag rate,
defined by the ratio of the number of tagged 
uds jets to the number of events in the initial
inclusive multihadronic event sample, is
$0.399\pm 0.001\,\mathrm{(stat.)}$
for the data and 
$0.418\pm 0.001\,\mathrm{(stat.)}$
for the Monte Carlo.
The difference of about 2\% between the
uds jet tag rates of data and Monte Carlo
implies a small deficiency in the simulation 
of the event characteristics,
which is accounted for in our evaluation
of systematic uncertainties
(section~\ref{sec-systematic}).
The corrected energy of the uds jets is given by
the beam energy, 45.6~GeV,
with essentially no uncertainty.

For purposes of comparison,
Figure~\ref{fig-mcnch91}
includes the prediction of Herwig for uds
{\qq} event hemispheres,
generated using the same c.m. energy
that is used to generate the {\gluglu} event sample.

\section{Corrections}
\label{sec-corrections}

The measured charged particle multiplicity distributions
of the {\gincl} and uds jets are corrected in two steps,
following the method presented in~\cite{bib-opalnch}.
In the first step,
the data are corrected
for experimental acceptance, resolution, 
and secondary electromagnetic and hadronic interactions
using an unfolding matrix~\cite{bib-opalnch}.
This matrix is constructed using Jetset events,
including full detector simulation and the same selection
criteria as are applied to the data.
In the second step,
the data are corrected for event acceptance and
the effects of initial-state photon radiation
using bin-by-bin multiplicative factors.
The bin-by-bin corrections are derived using 
two different Jetset samples.
The first sample,
based on inclusive Z$^0$ hadronic decays, 
includes initial-state photon radiation
and the same event acceptance criteria as the data,
but not detector simulation
(they are the generator level input to events
which have been processed through the detector simulation
and which have been selected using the same selection
criteria as are applied to the data).
The second sample does not include
initial-state photon radiation,
event acceptance,
or detector simulation
and treats all charged and neutral particles
with mean lifetimes greater than
\mbox{$3\times 10^{-10}$~s} as stable:
hence charged particles from the decays of K$_{\mathrm S}^0$
and weakly decaying hyperons are included 
in the definition of multiplicity.
For the correction of the gluon jet data,
inclusive Z$^0$ events are used for the second sample.
The quark jets in this sample are identified
with Monte Carlo information
using the method discussed in section~\ref{sec-gluon}:
otherwise the {\gincl} sample is obtained in the same
manner as is described in section~\ref{sec-gluon} for the data.
For the correction of the uds jet data,
the jets of the second sample are defined by the particles in
the hemispheres of uds events.
The multiplicative correction factors are obtained
by taking the ratios of the predictions from the second 
sample to those from the first one.
Thus, the bin-by-bin corrections account not only for 
detector response and initial-state radiation
but also for the background to the {\gincl} and
uds jet data.
The corrections applied to the data
are generally moderate or small.
For example,
Jetset predicts the mean multiplicity value
(section~\ref{sec-mean})
of {\gincl} jets to be only
6\% larger at the generator level
than it is at the level which includes
detector simulation and the
experimental selection criteria.
The corresponding difference for uds jets is~$-4$\%.

\section{Results}
\label{sec-results}

\begin{figure}[ht]
\epsfxsize=15cm
\epsffile[15   100 545 700]{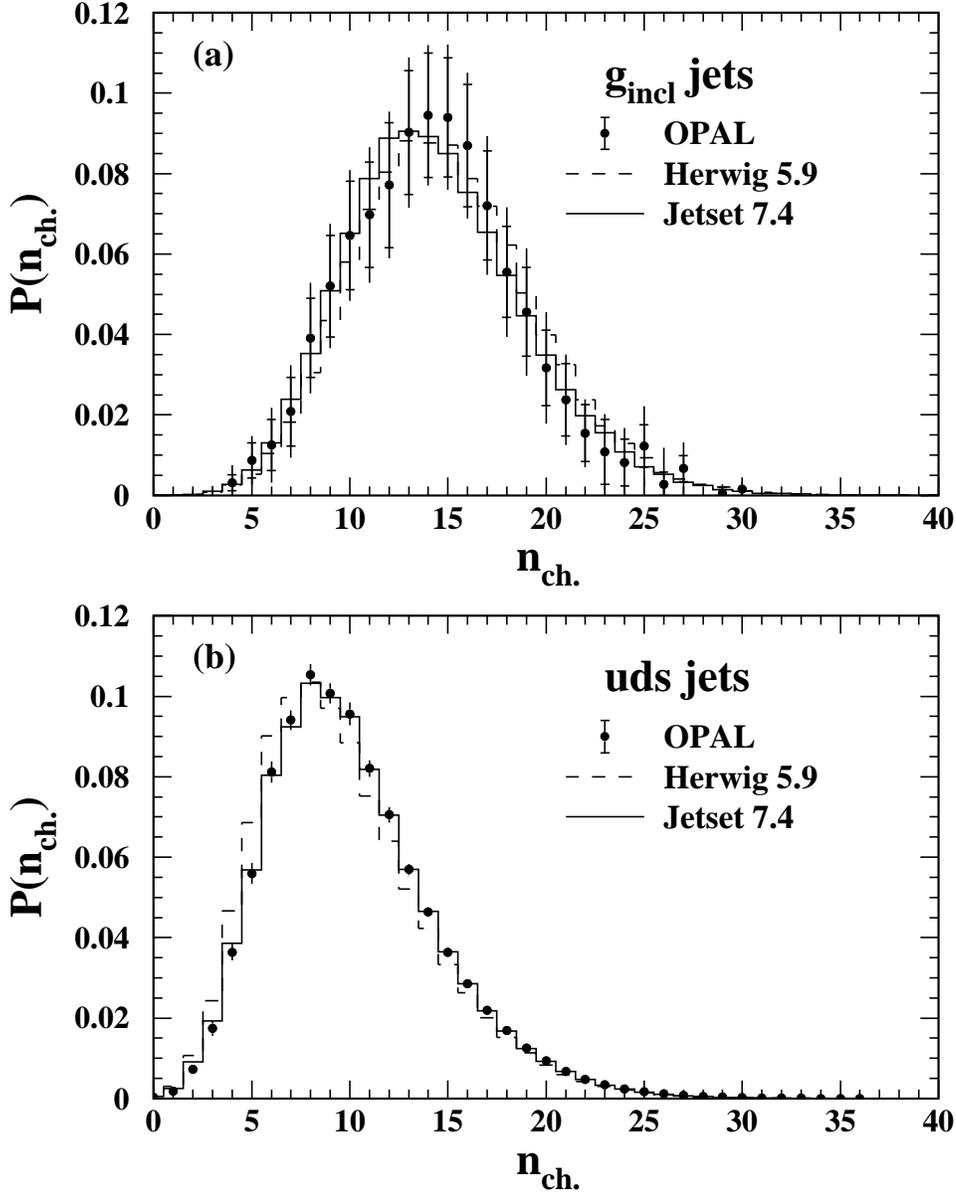}
\caption{
Corrected distributions of charged particle multiplicity
for (a)~41.8~GeV {\gincl} gluon jets,
and (b)~45.6~GeV uds quark jets.
The total uncertainties are shown by vertical lines.
The statistical uncertainties are
indicated by small horizontal bars.
(The statistical uncertainties are too small 
to be seen for the uds jets.)
The uncertainties are correlated between bins.
The predictions of the Herwig and Jetset
parton shower event generators are also shown.
Numerical values for these data are given in
Table~\ref{tab-nchresults}.
}
   \label{fig-nchresults}
\end{figure}

The corrected charged particle multiplicity distributions
are presented in Figure~\ref{fig-nchresults}.
Numerical values for these data
are listed in Table~\ref{tab-nchresults}.
Shown in comparison to the data are the generator
level predictions of Herwig~5.9 and Jetset~7.4.
The Monte Carlo results for the {\gincl} jets are
obtained in the manner described in
section~\ref{sec-ggcompare}.
The two models are seen to provide a generally
adequate description of the measurements
except that the uds jet distribution predicted by Herwig is
shifted towards lower values of multiplicity
than are observed experimentally
(Figure~\ref{fig-nchresults}(b)).
Statistical uncertainties were estimated for the
{\gincl} results
using 100 Monte Carlo samples of {\gincl} jets
at the generator level,
each with approximately the same event statistics
as the data sample.
The statistical uncertainty for each result
(e.g.~a multiplicity bin in Figure~\ref{fig-nchresults}(a)
or a factorial moment measurement,
see section~\ref{sec-moments})
was set equal to the RMS value found for
the 100~samples.
The same method was used to evaluate statistical
uncertainties for the uds jets.
The matrix corrections introduce correlations 
between the bins of the corrected
multiplicity distributions.
The correlations are generally strong between
a bin and its nearest one or two neighbors on either side
but can extend with smaller strength to
four or five bins away.
These correlations smooth out
bin-to-bin statistical fluctuations.
This effect is particularly noticeable for the gluon jet
distribution (Figure~\ref{fig-nchresults}(a))
because of the relatively small number of
events in the {\gincl} jet sample.

\subsection{Mean, dispersion, skew 
and curtosis values}
\label{sec-mean}

\begin{figure}[ht]
\epsfxsize=17cm
\epsffile[15   100 545 700]{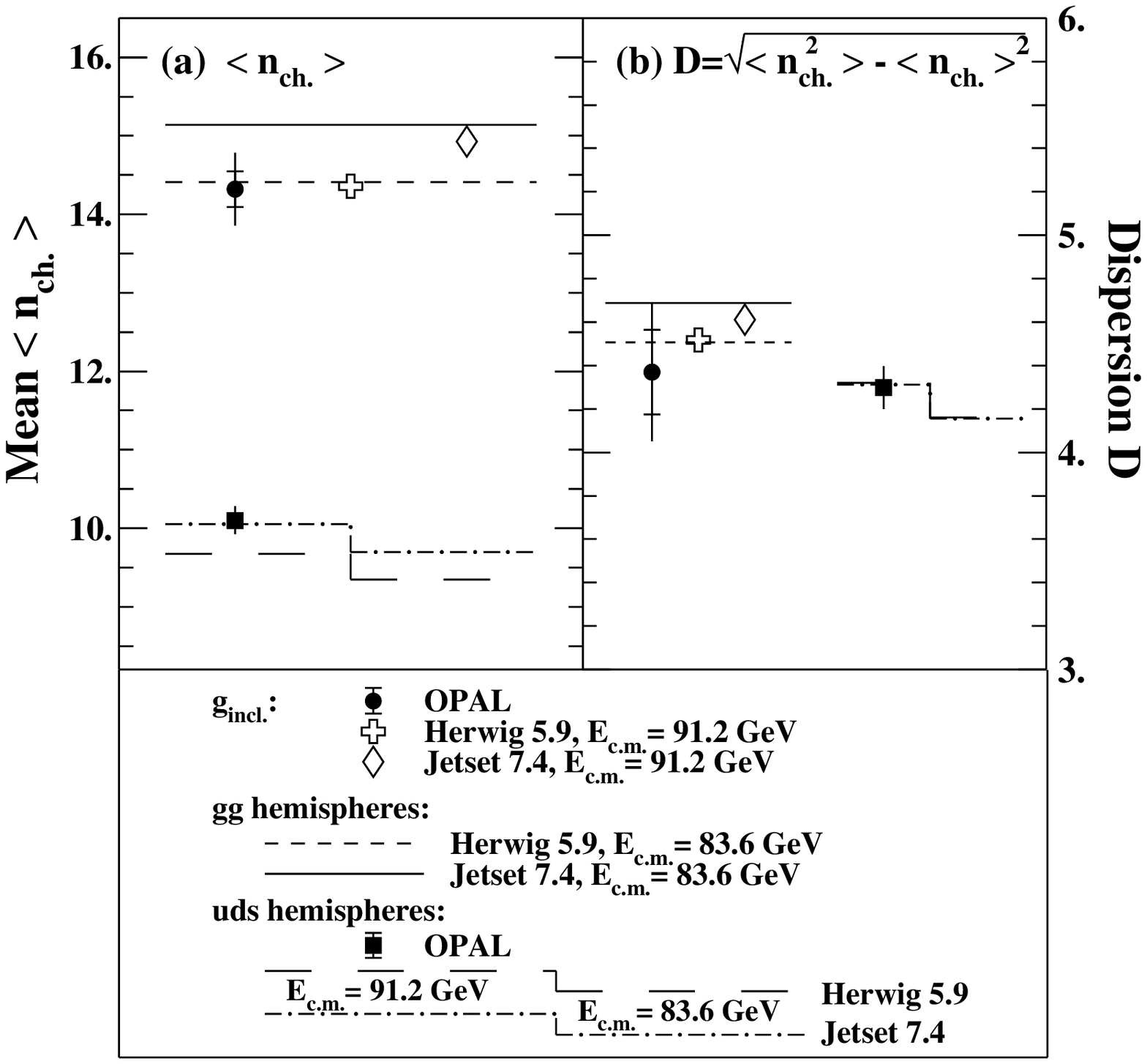}
\caption{
(a)~The mean, {\mnch},
and (b)~the dispersion, $D$,
of the charged particle 
multiplicity distribution of 
41.8~GeV {\gincl} gluon jets
and 45.6~GeV uds quark jets.
The total uncertainties are shown by vertical lines.
The statistical uncertainties are
indicated by small horizontal bars.
(The statistical uncertainties are too small 
to be seen for the uds jets.)
The predictions of the Herwig and Jetset
parton shower event generators are also shown. 
}
   \label{fig-mean-dispersion}
\end{figure}

\begin{figure}[ht]
\epsfxsize=17cm
\epsffile[15   100 545 700]{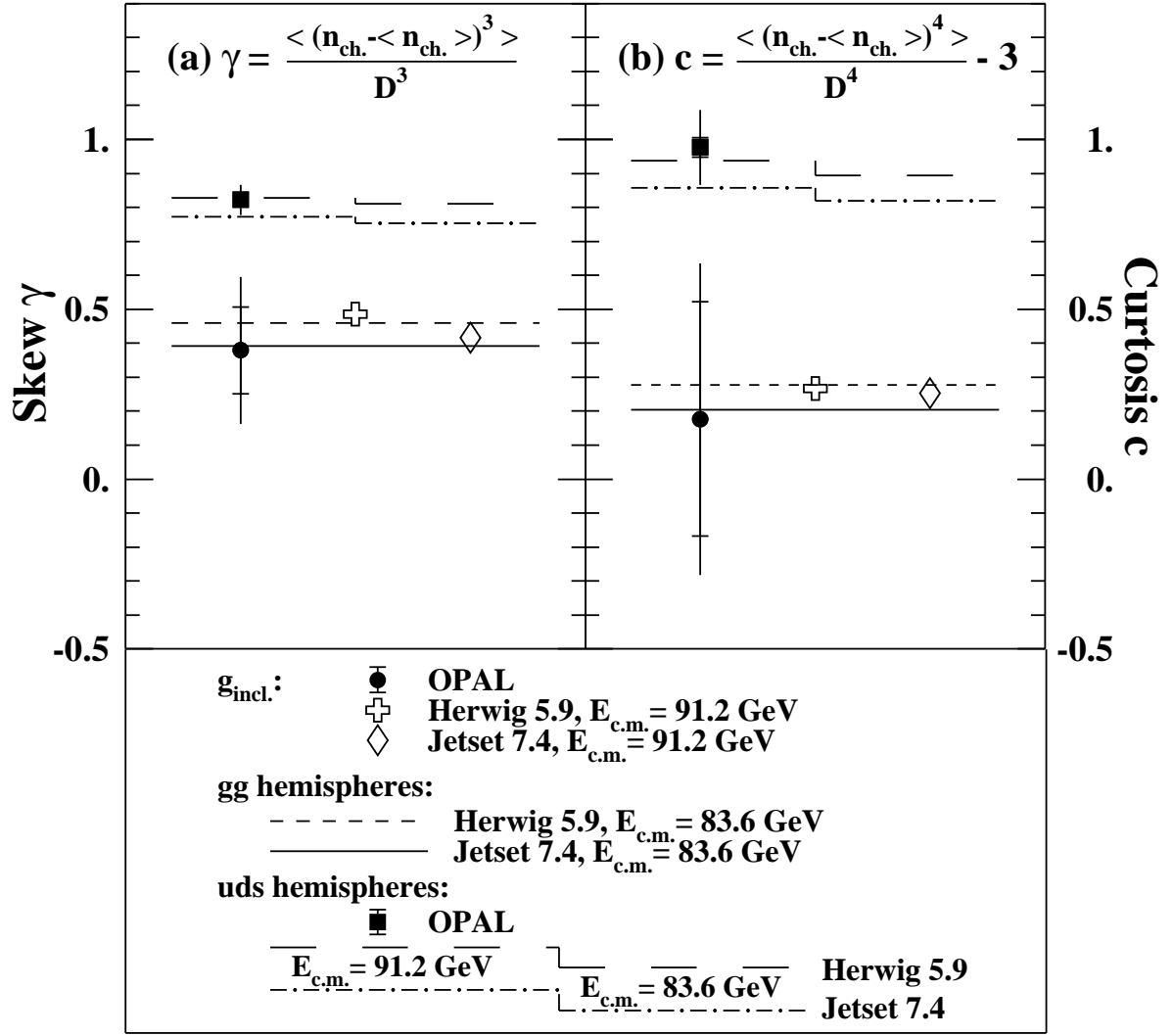}
\caption{
(a)~The skew, $\gamma$,
and (b)~the curtosis, $c$,
of the charged particle 
multiplicity distribution of 
41.8~GeV {\gincl} gluon jets
and 45.6~GeV uds quark jets.
The total uncertainties are shown by vertical lines.
The statistical uncertainties are
indicated by small horizontal bars.
(The statistical uncertainties are too small 
to be seen for the uds jets.)
The predictions of the Herwig and Jetset
parton shower event generators are also shown. 
}
   \label{fig-skew-curtosis}
\end{figure}

We determine the mean {\mnch}, dispersion
$D\equiv\sqrt{\langle n_{\,\mathrm{ch.}}^2 \rangle- 
\langle n_{\,\mathrm{ch.}} \rangle^2}$,
skew $\gamma\equiv\langle(n_{\,\mathrm{ch.}}-
\langle n_{\,\mathrm{ch.}}\rangle)^3\rangle/D^3$
and curtosis
$c\equiv[(\langle(n_{\,\mathrm{ch.}}-
\langle n_{\,\mathrm{ch.}}\rangle)^4\rangle/D^4)-3]$
values of 41.8~GeV {\gincl} gluon jet and
45.6~GeV uds quark jet hemispheres to be:
\begin{eqnarray}
  \label{eq-results}
   \langle n_{\,\mathrm{ch.}} 
      \rangle_{\mathrm{g}_{\,\mathrm{incl.}}} & = &  
      14.32\pm0.23\pm0.40  \\
   \langle n_{\,\mathrm{ch.}} 
      \rangle_{\mathrm{uds\,hemis.}} & = &
      10.10\pm0.01\pm0.18 \nonumber \\[2mm]
   D_{\mathrm{g}_{\,\mathrm{incl.}}} & = & 
      4.37\pm0.19\pm0.26  \\
   D_{\mathrm{uds\,hemis.}} & = &
      4.298\pm0.008\pm0.098 \nonumber \\[2mm]
   \gamma_{\mathrm{g}_{\,\mathrm{incl.}}} & = & 
      0.38\pm0.13\pm0.18 \\
   \gamma_{\mathrm{uds\,hemis.}} & = &
      0.822\pm0.007\pm0.044 \nonumber \\[2mm]
   c_{\,\mathrm{g}_{\,\mathrm{incl.}}} & = & 
      0.18\pm0.34\pm0.30 \\
   c_{\,\mathrm{uds\,hemis.}} & = &
      0.98\pm0.03\pm0.11 \nonumber \;\;\;\; ,
\end{eqnarray}
where the first uncertainty is statistical and
the second is systematic.
These results are shown in 
Figures~\ref{fig-mean-dispersion}
and~\ref{fig-skew-curtosis}.
The systematic uncertainties are discussed below
in section~\ref{sec-systematic}.
The results for {\mngincl} and {\mnuds} are
consistent with those presented 
in~\cite{bib-opalqg96,bib-opalsll}.

Figures~\ref{fig-mean-dispersion}
and~\ref{fig-skew-curtosis}
include the Herwig and Jetset predictions for {\gincl} jets,
shown by the cross and diamond symbols, respectively.
Also shown, by the finely-dashed and solid horizontal lines,
are the Monte Carlo predictions for {\gluglu} event hemispheres,
generated to have the same energy as the {\gincl} jets.
It is seen that the results for the {\gluglu} hemispheres
and {\gincl} samples agree well for both Herwig and Jetset
(compare the cross symbols to the finely-dashed lines and
the diamond symbols to the solid lines),
which is consistent with Figure~\ref{fig-mcnch91}
and the discussion in section~\ref{sec-ggcompare}.
The Monte Carlo predictions for gluon jet properties
are seen to agree well with the data,
except that the Jetset prediction of
the mean multiplicity {\mngincl}
is somewhat above the measured value
(Figure~\ref{fig-mean-dispersion}(a)).

Also shown in 
Figures~\ref{fig-mean-dispersion}
and~\ref{fig-skew-curtosis},
by the coarsely-dashed and dash-dotted horizontal lines,
are the predictions of Herwig and Jetset for uds event hemispheres.
These predictions are shown for two different values of jet energy:
{\ejet}=45.6~GeV ({\ecm}=91.2~GeV),
corresponding to the energy of the measured uds jets,
and {\ejet}=41.8~GeV ({\ecm}=83.6~GeV), 
corresponding to the energy of the {\gincl} jets.
The steps in 
Figures~\ref{fig-mean-dispersion}
and~\ref{fig-skew-curtosis}
between the predictions
for {\ecm}=91.2~GeV and {\ecm}=83.6~GeV uds hemispheres
therefore indicate the Monte Carlo corrections for quark jets
to account for the difference in energy between the
uds and {\gincl} samples.
Comparing the Monte Carlo predictions for {\ecm}=91.2~GeV
to the data,
it is seen that Jetset provides a good overall 
description of the uds jet properties.
Herwig's predictions for the uds jet properties
are also in reasonable agreement with the data,
with the exception of the mean multiplicity
{\mnuds} (Figure~\ref{fig-mean-dispersion}(a)),
which is somewhat too low as was already noted
in connection with Figure~\ref{fig-nchresults}(b).

We also determine the ratios between
the gluon and quark jet results since common
systematic uncertainties will partially cancel.
Before forming these ratios,
it is necessary to account for the different
energies of the two samples:
the gluon jets have a mean energy of 41.8~GeV while
the uds jets have a mean energy of 45.6~GeV.
To correct the quark jet {\mnch} value for this
difference in energy,
we follow the method in~\cite{bib-opalqg96}
and employ the QCD analytic formula for the evolution
of the mean event multiplicity in {\epem} 
annihilations~\cite{bib-qcdnecm}.
This QCD result is known to describe the
energy evolution of the mean charged particle
multiplicity in inclusive {\epem} annihilation events
with good accuracy~\cite{bib-opallep2}.
Assuming the number of active quark flavors, $n_f$,
to be five,
the QCD evolution formula predicts the mean multiplicity
of 41.8~GeV quark jet hemispheres to be ($3.6\pm 0.2$)\% 
smaller than for 45.6~GeV quark jet hemispheres,
where the uncertainty results from the maximum
variation found by
using the jet energies (41.8~GeV and 45.6~GeV)
rather than the event energies (83.6~GeV and 91.2~GeV),
$n_f$=3 rather than $n_f$=5,
and varying the value of {\lms} within its allowed
range~\cite{bib-pdg96}.\footnote{Jetset and Herwig
predict reductions of 3.5\% and 3.6\%,
respectively, in the value of
{\mnch} for uds hemispheres with {\ecm}=83.6~GeV
compared to those with {\ecm}=91.2~GeV:
thus the Monte Carlo 
predictions for the energy correction
are essentially the same as that obtained from the QCD
evolution formula.}
Virtually the same result is obtained if the evolution
formula is evaluated using the fitted values 
given in~\cite{bib-opallep1p5}
for the strong coupling strength
and the overall normalization.
Applying a multiplicative correction of \mbox{$0.964$}
to the result presented above for {\mnuds} yields
{\mnudsmod}=$9.74\pm 0.01\,\mathrm{(stat.)}$.
Our result for the multiplicity ratio {\rch} between
41.8~GeV gluon and quark jets is therefore:
\begin{equation}
\label{eq-rch}
   r_{\,\mathrm{ch.}} \equiv
   \frac{ \langle n_{\,\mathrm{ch.}} 
      \rangle_{\mathrm{g}_{\,\mathrm{incl.}}} }
   { \langle n_{\,\mathrm{ch.}} 
      \rangle_{\mathrm{uds\,\,hemis.}}^{41.8\,\mathrm{GeV}} }
   =
   1.471\pm 0.024\,\mathrm{(stat.)}\pm 0.043\,\mathrm{(syst.)}
      \;\;\;\; .
\end{equation}
This result is consistent with our
previous result~\cite{bib-opalqg96},\footnote{There
is a shift of 0.081 between
the central values of our previous result
({\rch}=$1.552\pm0.073$~(stat.+syst.))
and our current result;
this shift is due primarily to our reevaluation
of the {\gincl} energy value (section~\ref{sec-gluon})
and thus to a reevaluation of
the correction to account for the difference
in energy between the uds and {\gincl} jets.}
but has substantially reduced uncertainties.
Furthermore,
the analytic prediction in~\cite{bib-qcdmult3}
is in general agreement with this measurement.
For purposes of comparison,
the predictions of Herwig and Jetset are
$r_{\,\mathrm{ch.}}$$=$$1.537\pm0.002$ 
and $1.539\pm0.002$, respectively,
where the uncertainties are statistical.

For the dispersion, skew and curtosis values,
we account for the difference
in energy between the gluon and quark jet measurements
using the Monte Carlo predictions.
The Monte Carlo is known to provide a good
description of the energy evolution of the
dispersion of the multiplicity distribution
in inclusive {\epem} hadronic events
(e.g.~see~\cite{bib-opalnch,bib-opallep2,bib-opallep1p5}),
making it plausible that its predictions for the
energy evolution of skew and curtosis
are also reliable.
Jetset predicts the dispersion $D$ 
to be $(3.6\pm0.2)$\% smaller
for 41.8~GeV uds hemispheres than for 
45.6~GeV uds hemispheres 
(Figure~\ref{fig-mean-dispersion}(b)).
The corresponding results for skew and curtosis
are~$(2.5\pm0.5)$\% and $(4.5\pm2.2)$\%, respectively
(Figure~\ref{fig-skew-curtosis}(a) and~(b)).
The uncertainties for these values are 
given by the maximum variation found by
using Herwig rather than Jetset,
$n_f$=5 rather than $n_f$=3,
and varying the value of $\Lambda_{\mathrm{LLA}}$
by its uncertainty~\cite{bib-qg95b}.
Applying corrections of $0.964$, $0.975$ 
and $0.955$ to the
uds jet dispersion, skew and curtosis measurements 
given above, respectively, yields
$D_{\,\mathrm{uds\,hemis.}}^{\,41.8\,\mathrm{GeV}} =
   4.143\pm 0.008\,\mathrm{(stat.)}$,
$\gamma_{\,\mathrm{uds\,hemis.}}^{\,41.8\,\mathrm{GeV}} =
   0.802\pm 0.007\,\mathrm{(stat.)}$, and
$c_{\,\mathrm{uds\,hemis.}}^{\,41.8\,\mathrm{GeV}} =
   0.933\pm 0.027\,\mathrm{(stat.)}$.
The ratios between the dispersion, 
skew and curtosis values of 41.8~GeV gluon
and quark jets are therefore:
\begin{eqnarray}
\label{eq-disp}
   r_{D} \equiv
    \frac{ D_{\mathrm{g}_{\,\mathrm{incl.}}} }
   { D_{\,\mathrm{uds\,\,hemis.}}^{\,41.8\,\mathrm{GeV}} }
   & = &  
   1.055\pm 0.046\,\mathrm{(stat.)}\pm 0.055\,\mathrm{(syst.)}
   \\
   r_{\gamma} \equiv
   \frac{ \gamma_{\mathrm{g}_{\,\mathrm{incl.}}} }
   { \gamma_{\,\mathrm{uds\,\,hemis.}}^{\,41.8\,\mathrm{GeV}} }
   & = &  
   0.47\pm 0.16\,\mathrm{(stat.)}\pm 0.21\,\mathrm{(syst.)}
   \\
   r_{c} \equiv
   \frac{ c_{\,\mathrm{g}_{\,\mathrm{incl.}}} }
   { c_{\,\mathrm{uds\,\,hemis.}}^{\,41.8\,\mathrm{GeV}} }
   & = &  
   0.19\pm 0.37\,\mathrm{(stat.)}\pm 0.33\,\mathrm{(syst.)}
   \;\;\;\; .
\end{eqnarray}
The dispersions of the gluon and quark jet multiplicity 
distributions are therefore almost equal,
despite the large difference between their
mean values (relation~(\ref{eq-rch})).
The quark jet distribution 
is more skewed,
i.e.~asymmetric,
than the corresponding distribution
for gluon jets.
Quark jets are observed to have a larger
curtosis value than gluon jets,
implying that their distribution is
more non-gaussian in shape
(the peak is higher and the tails are broader than 
a gaussian with the same mean and dispersion)
than is the case for gluon jets.
We note that the deviations of the ratios
$r_{\gamma}$ and $r_{c}$
from unity are only 2.0 and 1.6~standard deviations 
of their total uncertainties, however.
Herwig predicts
$r_{D}$$=$$1.086\pm0.003$, $r_{\gamma}$$=$$0.60\pm0.01$
and $r_{c}$$=$$0.30\pm0.03$,
where the uncertainties are statistical.
The corresponding results from Jetset
are $1.110\pm0.003$, $0.55\pm0.01$ and $0.31\pm0.03$.

\subsection{Factorial, cumulant and \boldmath{$\hq$} moments}
\label{sec-moments}

Factorial moments provide a standard means to
characterize the fluctuations of a distribution
about its mean value
(cf.~\cite{bib-webber,bib-analytic1,bib-bialas}).
Factorial moments are less subject to bias from
random statistical fluctuations than
``ordinary'' central moments,
as is discussed in~\cite{bib-bialas}.
Various QCD analytic calculations have been performed
for the factorial moments of the multiplicity
distributions of separated gluon and quark
jets~\cite{bib-webber,bib-analytic1}.
A factorial moment analysis of our data
permits a test of these QCD calculations
for the first time (see section~\ref{sec-analytic}).
Currently, QCD predictions do not exist for
the dispersion, skew and curtosis values
presented in section~\ref{sec-mean}.
Certain combinations of factorial moments are 
directly related to these three quantities,
however, as is discussed below.

The normalized factorial moment of rank $q$, {$\fq$},
is defined by~\cite{bib-bialas,bib-dreminrev}:
\begin{equation}
  \fq\equiv\displaystyle{
    \frac{ \langle n\,(n-1)\cdots(n-q+1)\rangle }
    { \langle n \rangle^q }
    }    \;\;\;\; ,
\end{equation}
where \mbox{$q\geq 1$}.
An equivalent characterization is given by the
cumulant factorial moments, {$\kq$}~\cite{bib-dreminrev},
which can be obtained from the {$\fq$} moments recursively:
\begin{equation}
  \kq\equiv\fq - \displaystyle{\sum_{m=1}^{q-1}
    \frac{ (q-1)! }{ m! (q-m-1)! }
    \,K_{q-m}\,F_m    \;\;\;\; ,
   }
\end{equation}
with the condition $K_1$=1.
The {$\kq$} moments have been shown to be more
sensitive to detailed features of the multiplicity
distribution than the {$\fq$} moments~\cite{bib-dreminrev}:
they are more sensitive to higher order QCD corrections
and to differences between QCD and
phenomenological parametrizations such
as the negative binomial distribution.
Besides factorial and cumulant moments,
it has become standard to consider the
ratio of cumulant to factorial moments,
denoted {$\hq$},
which appear naturally in the solution
of the QCD equations~\cite{bib-dreminpriv}:
\begin{equation}
    \hq \equiv \frac{ \kq }{ \fq }  \;\;\;\; .
\end{equation}
An analysis of the {$\hq$} factorial moments of the
inclusive multiplicity distribution in
multihadronic Z$^0$ decays has recently been presented 
in~\cite{bib-sldhq,bib-hqothers}.

\begin{figure}[ht]
\epsfxsize=15cm
\epsffile[15   100 545 700]{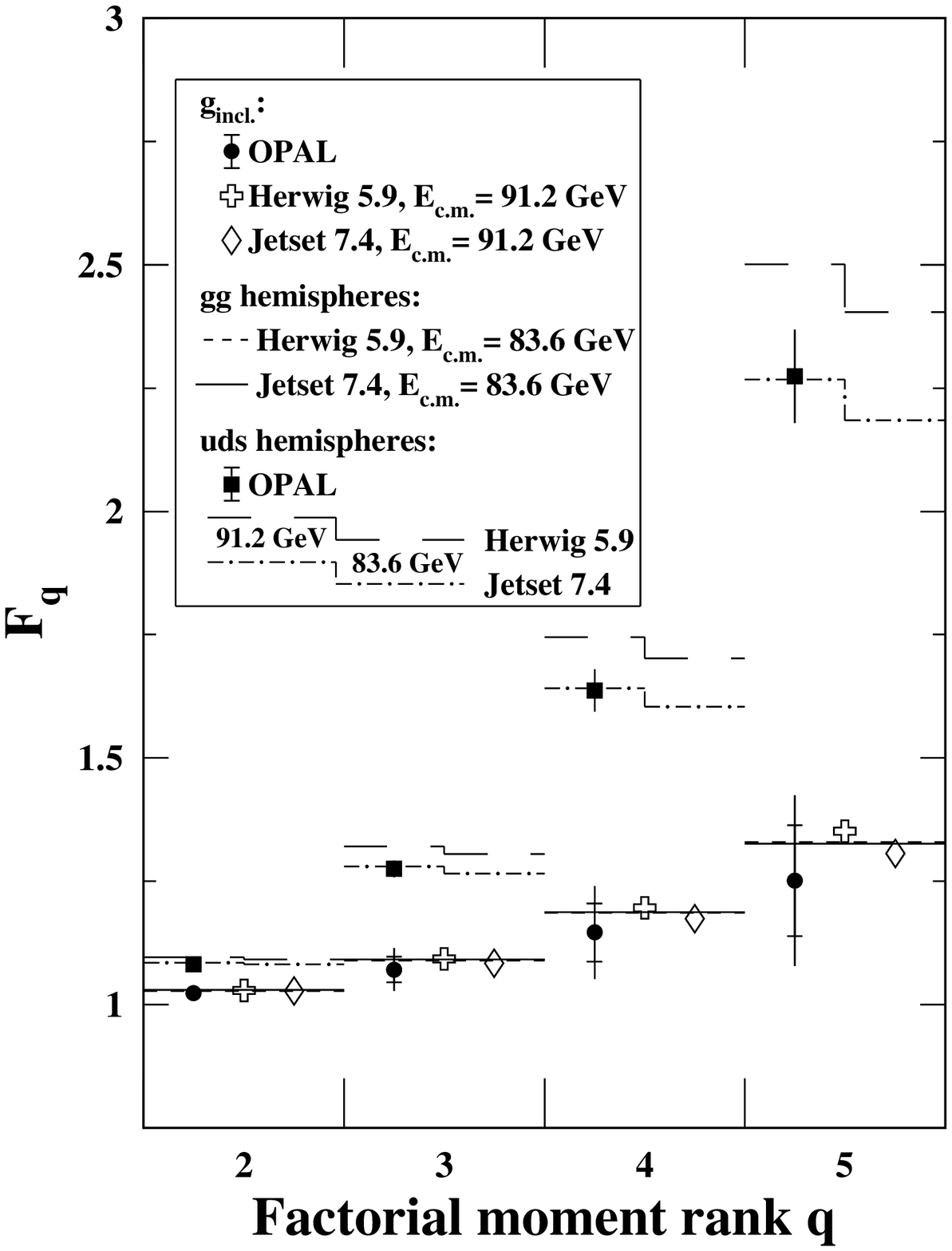}
\caption{
The normalized factorial moments of the charged particle
multiplicity distribution, {$\fq$},
for 41.8~GeV {\gincl} gluon jets
and 45.6~GeV uds quark jets.
The total uncertainties are shown by vertical lines.
The statistical uncertainties are
indicated by small horizontal bars.
(The statistical uncertainties are too small 
to be seen for the uds jets.)
The predictions of the Herwig and Jetset
parton shower event generators are also shown. 
}
   \label{fig-fqlow}
\end{figure}

\begin{figure}[ht]
\epsfxsize=15cm
\epsffile[15   100 545 700]{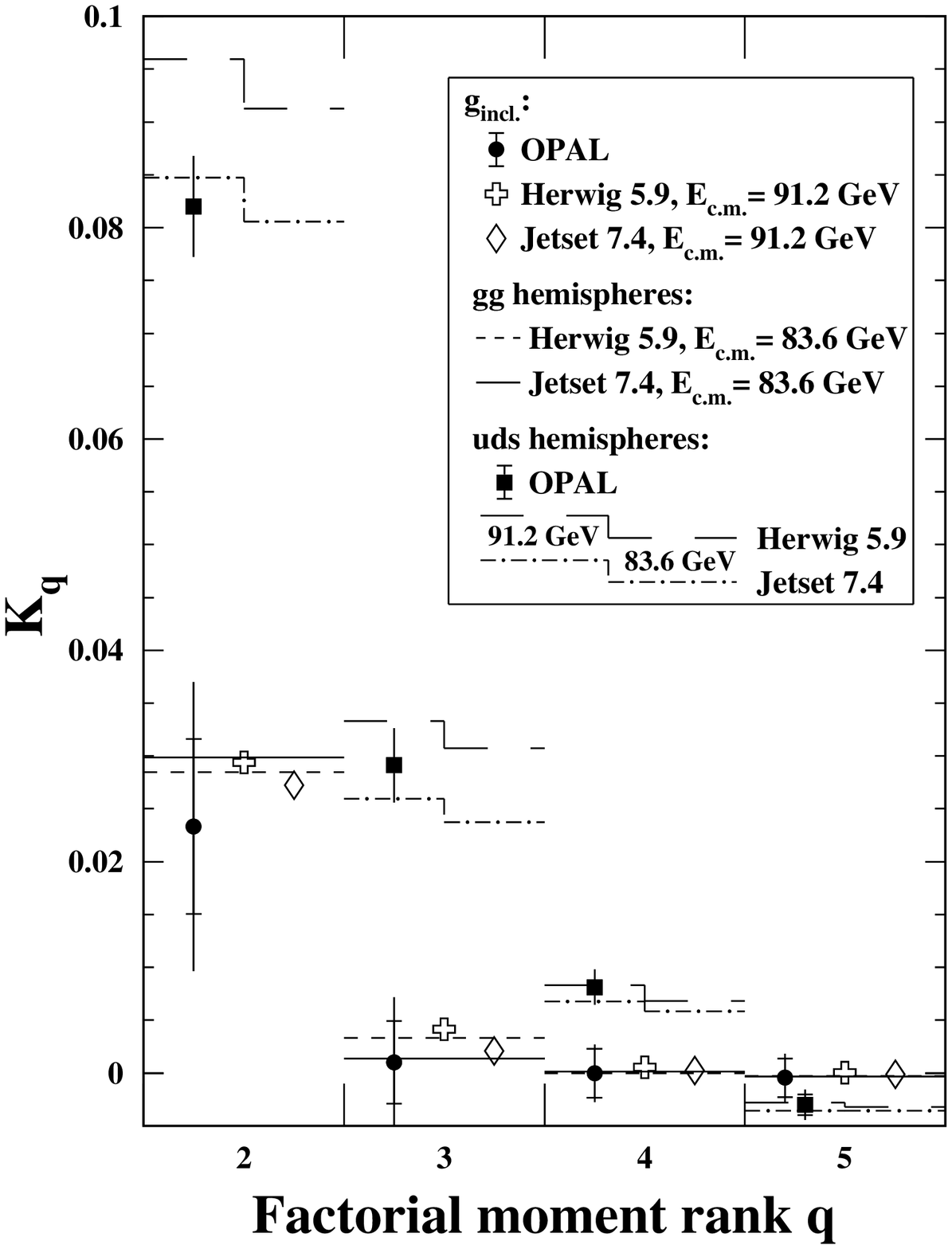}
\caption{
The cumulant factorial moments of the charged particle
multiplicity distribution, {$\kq$},
for 41.8~GeV {\gincl} gluon jets
and 45.6~GeV uds quark jets.
The total uncertainties are shown by vertical lines.
The statistical uncertainties are
indicated by small horizontal bars.
(The statistical uncertainties are too small 
to be seen for the uds jets.)
The predictions of the Herwig and Jetset
parton shower event generators are also shown. 
}
   \label{fig-kqlow}
\end{figure}

In Tables~\ref{tab-gmoments} and~\ref{tab-qmoments},
we present our measurements of the 
{$\fq$}, {$\kq$} and {$\hq$} factorial moments
of {\gincl} and uds jets.
Our results are given for ranks \mbox{$2\leq q\leq 5$}.
($F_1$=$K_1$=$H_1$=1 trivially.)
These ranks correspond to those for which theoretical
predictions have been published~\cite{bib-webber,bib-analytic1}.
The results for {$\fq$} and {$\kq$} are
shown in Figures~\ref{fig-fqlow} and~\ref{fig-kqlow}.
The results for {$\hq$} are qualitatively
similar to those shown in Figure~\ref{fig-kqlow}
for {$\kq$} and so are not shown in addition.
Figures~\ref{fig-fqlow} and~\ref{fig-kqlow} include
the predictions of Herwig and Jetset.
The Monte Carlo information
is presented in the same manner as in 
Figures~\ref{fig-mean-dispersion}
and~\ref{fig-skew-curtosis}.
It is seen that the Monte Carlo results for the {\gincl}
and {\gluglu} event hemispheres
agree well with each other,
i.e. the cross symbols agree with the finely-dashed lines
and the diamond symbols agree with the
solid lines to within differences that are
consistent with statistical fluctuations.
It is also seen that the Monte Carlo corrections to uds jets
to account for the difference in energy between 
the {\gincl} and uds samples
(the steps in the center of the coarsely-dashed
and dash-dotted curves in 
Figures~\ref{fig-fqlow} and~\ref{fig-kqlow})
are moderate in comparison
to the experimental uncertainties.

To gain insight concerning the physical interpretation
of the factorial and cumulant moments,
and to help relate the measurements shown
in Figures~\ref{fig-fqlow} and~\ref{fig-kqlow}
to those shown in
Figures~\ref{fig-mean-dispersion}
and~\ref{fig-skew-curtosis},
we generated 100 samples of {\gincl} jets using
the Jetset Monte Carlo,
each with event statistics similar to that of the data.
We used these 100 generator level
samples to calculate the correlation
coefficients between the {$\fq$} moments,
the {$\kq$} moments
and the mean, dispersion, skew and curtosis values.
The resulting correlation matrix
is presented in Table~\ref{tab-fqcorr}.
Similar results were found in an analogous
study of uds quark jets.
We also selected 24 independent samples of uds
events which included detector simulation and the
same selection criteria as the data.
We processed these 24 samples using the correction
procedure described in section~\ref{sec-corrections}
and determined the correlations between the
corrected results.
The detector level study was not repeated for {\gincl}
jets because of inadequate Monte Carlo event statistics.
The resulting correlation matrix
(corresponding to Table~\ref{tab-fqcorr})
was found to be very similar to that obtained
using the 100 generator level samples,
from which we conclude that the correction procedure
does not introduce significant correlations between
the variables.
To facilitate the discussion in the next paragraph,
boxes have been drawn around correlation
coefficients in Table~\ref{tab-fqcorr}
which have magnitudes of~0.90 or larger.

From Table~\ref{tab-fqcorr},
it is seen that there is a high degree of 
statistical correlation between factorial
moments of different rank:
the correlation coefficients between the various 
{$\fq$} moments lie between 0.92 and~0.99.
In contrast,
only modest or small correlations are observed
between the cumulant moments.
The largest correlation coefficient in this case,
between the {$\kq$} moments with
ranks $q$=3 and $q$=4, is only 0.43.
Thus, the results shown for different ranks 
in Figure~\ref{fig-kqlow}
are largely independent of each other,
while those shown in Figure~\ref{fig-fqlow} are not.
Table~\ref{tab-fqcorr} also establishes that
the cumulant moments of ranks 2, 3 and~4 are strongly
correlated with dispersion, skew and curtosis, respectively
(correlation coefficients of 0.97, 0.99 and~0.90).
Furthermore,
with the exception of the correlation between
$K_3$ and curtosis (coefficient of~0.78),
the other correlations of the cumulant moments with 
dispersion, skew and curtosis are moderate or small.
Therefore, the cumulant moments of ranks 2, 3 and 4
are directly related to
dispersion, skew and curtosis, respectively.
Algebraically,
the relationships are:
\begin{eqnarray}
\label{eq-k2}
  K_2 & = & 
      \left( \frac{D}{\mnchmath}\right)^2 
        - \;\;\frac{1}{\mnchmath}  \\
  K_3 & = &
      \gamma\;\left( \frac{D}{\mnchmath}\right)^3
      -\;\;\frac{3}{\mnchmath}
       \left( \frac{D}{\mnchmath}\right)^2
      +\;\;\frac{2}{{\mnchmath}^2}  \\
\label{eq-k4}
  K_4 & = &
      c\;\left( \frac{D}{\mnchmath}\right)^4
      -\;\;\frac{6\gamma}{\mnchmath}
       \left( \frac{D}{\mnchmath}\right)^3
      +\;\;\frac{11}{{\mnchmath}^2}
       \left( \frac{D}{\mnchmath}\right)^2
      -\;\;\frac{6}{{\mnchmath}^3}
   \;\;\;\; .
\end{eqnarray}
Thus, the {$\kq$} moments with $q$=2, 3 and~4 
are essentially equivalent
to dispersion, skew and curtosis,
but are in a form for which QCD analytic
calculations have been presented
(see section~\ref{sec-analytic}).
In contrast,
the {$\fq$} moments exhibit a strong correlation
with dispersion but not with skew or curtosis,
as is seen from Table~\ref{tab-fqcorr}.

In Table~\ref{tab-rmoments},
we present measurements of the ratios
{$\rfq$}, {$\rkq$} and {$\rhq$}
between the {$\fq$}, {$\kq$} and {$\hq$} factorial moments 
of 41.8~GeV gluon and quark jets.
To obtain these results,
the quark jet values in Table~\ref{tab-qmoments}
were corrected for the difference in energy between
the uds and {\gincl} samples using the method 
described in section~\ref{sec-mean}
for the dispersion, skew and curtosis,
i.e.~using the Jetset predictions
(e.g.~Figures~\ref{fig-fqlow} and~\ref{fig-kqlow}),
with a systematic uncertainty evaluated as
is described in section~\ref{sec-mean}.
These corrections typically lie between 0.95 and~0.99.
The ratios {$\rfq$}, {$\rkq$} and {$\rhq$} are 
then formed by dividing the factorial moments of 
gluon jets (Table~\ref{tab-gmoments})
by these corrected quark jet results.
Our measurements of the cumulant moment ratios 
{$\rkq$} are shown in Figure~\ref{fig-kqratio}.
It is seen that the gluon and quark jet cumulant 
moments differ by about a factor of three for 
$q$=2 and by an even larger amount for the 
higher moments.
From relations~(\ref{eq-k2})-(\ref{eq-k4}),
it is seen that part of this difference can be
attributed to the difference between the mean 
values~{\mnch} of gluon and quark jets
(relation~(\ref{eq-rch})).
The results are well reproduced by 
the predictions of Herwig and Jetset,
shown by the dashed and solid horizontal lines
in Figure~\ref{fig-kqratio}.

\subsection{Systematic Uncertainties}
\label{sec-systematic}

To evaluate systematic uncertainties,
the analysis was repeated with the
following changes relative to the standard analysis.
There were no significant changes in the number
of selected events or in their estimated purities 
compared to the standard results
unless otherwise noted.
\begin{enumerate}
\item  Charged tracks alone were used for the data and
  for the Monte Carlo samples which include detector simulation,
  rather than charged tracks plus unassociated
  electromagnetic clusters.
\item  Herwig was used to determine the correction matrix
  and bin-by-bin correction factors, rather than Jetset.
\item  The particle selection was varied, 
  first by restricting
  charged tracks and electromagnetic clusters
  to the central region of the detector,
  $|\cos(\theta)|<0.70$,
  rather than $|\cos(\theta)|<0.94$
  for the charged tracks and
  $|\cos(\theta)|<0.98$ for the clusters,
  and second by increasing
  the minimum momentum of charged tracks, \pminch,
  from 0.10~GeV/$c$ to 0.20~GeV/$c$.
\item  The gluon jet selection was performed using the
  JADE-E0~\cite{bib-jade} and cone~\cite{bib-opalcone}
  jet finders to define the tagged quark jets,
  rather than the {\durham} jet finder:
  320 and 246 {\gincl} jets resulted, respectively,
  of which 88\% and 76\% were in common with the
  events of the standard {\gincl} sample.
\item  The geometric conditions for the
  gluon jet selection were varied, 
  first by requiring the angle between the two
  jets in the tagged hemisphere to exceed 65$^\circ$,
  rather than~50$^\circ$,
  and second by requiring the two tagged quark jets
  to lie within 65$^\circ$ of the thrust axis,
  rather than~70$^\circ$.
\item At least one track with a signed impact parameter
  significance greater than~2.5 was required to be present in
  the displaced secondary vertices used to tag quark
  jets for the {\gincl} identification,
  rather than at least two;
  the {\gincl} sample increased to $1127$ jets,
  while its estimated gluon jet purity decreased to~55.8\%.
\item  The gluon jet sample was restricted to events 
  collected within 100~MeV of the Z$^0$ peak.
\item  uds jets were tagged using charged tracks
  that appeared within a cone of half angle 70$^\circ$
  around the thrust axis, rather than~40$^\circ$.
\item The maximum signed impact parameter significance
  of tracks used for the identification of uds jets was 
  increased from 1.5 to~2.5.
\item  For the ratios of mean multiplicity, dispersion,
  skew and curtosis
  {\rch}, $r_D$, $r_{\gamma}$ and $r_{\,c}$,
  and for the ratios of factorial moments
  {$\rfq$}, {$\rkq$} and {$\rhq$},
  the energy to which the quark jet results were
  corrected was varied by the total uncertainty
  of the {\gincl} jet energy (section~\ref{sec-gluon});
  also, for these same quantities,
  the correction factors to account for the difference
  between the uds and {\gincl} jet energies were
  varied by their uncertainties
  (sections~\ref{sec-mean} and~\ref{sec-moments}).
\end{enumerate}
The differences between the standard results
and those found using each of these conditions
were used to define symmetric systematic uncertainties.
For items~3, 4, 5 and~10,
the larger of the two described differences
with respect to the standard result
was assigned as the systematic uncertainty.
For item~2,
the difference with respect to the standard result
was multiplied by 2/$\sqrt{12}$~\cite{bib-barlow}
since Herwig represents an extreme choice of
hadronization model compared to Jetset.
For the uds jet differential multiplicity distribution
(Figure~\ref{fig-nchresults}(b)
and Table~\ref{tab-nchresults}),
we evaluated the systematic terms 
involving~{\pminch} (item~3 in the above list)
using the procedure described in~\cite{bib-opalnch}:
the corrected distribution was parameterized
using polynomials,
the parameterized distribution was shifted
along the multiplicity axis so that its mean
coincided with the mean of the standard result,
and the systematic uncertainty was defined
bin-by-bin by the difference between the
shifted and standard distributions.

The uncertainties were added in quadrature
to define the total systematic uncertainty.
For the differential multiplicity distributions
(Figure~\ref{fig-nchresults}
and Table~\ref{tab-nchresults}),
the systematic uncertainty evaluated for each bin 
was averaged with the results from its two neighbors
to reduce the effect of bin-to-bin fluctuations
(the single neighbor was used for 
bins on the endpoints of the distributions).
The largest systematic terms for the {\gincl} jet 
measurements were generally found to arise about equally
from items~3, 4 and~5 in the above list.
The largest systematic terms for the uds jet measurements
were generally found to arise from item~3 and, 
to a lesser extent, from item~1.
For the ratios of the gluon to quark jet measurements,
the largest systematic terms generally arose
from items~1 and~3-5.
As an illustration,
Table~\ref{tab-rchsyst} provides
a breakdown of the systematic uncertainties
evaluated for {\mngincl}, {\mnuds}, and~{\rch}.

From Table~\ref{tab-rchsyst},
it is seen that using the JADE-E0 or cone jet finders 
to identify the tagged quark jets for the {\gincl} sample
(item~4 in the above list), yields
{\rch}=$1.488\pm0.024$~(stat.) or
{\rch}=$1.448\pm0.028$~(stat.), respectively,
which differ by less than 2\% from
the standard result of
{\rch}=$1.471\pm0.024$~(stat.).
This emphasizes that our results are almost
independent of the choice of the jet finding algorithm.
In contrast,
results for {\rch} based on exclusive samples of three-jet 
{\qq}g events vary from 
{\rch}$=$$1.10\pm0.03\,
{\mathrm{(stat.+syst.)}}$~\cite{bib-qg95a}
for the cone jet finder to 
{\rch}$=$$1.37\pm0.04\,
{\mathrm{(stat.+syst.)}}$~\cite{bib-delqg96}
for the JADE-E0 jet finder and thus exhibit a 
strong dependence on the jet algorithm employed.

\section{Tests of QCD analytic predictions}
\label{sec-analytic}

A number of QCD analytic calculations have been presented for
the factorial moments of the multiplicity distributions
of separated gluon and quark jets.
Our data permit a test of these calculations
for the first time.
In the following,
we test the predictions of analytic calculations
for the cumulant moments~{$\kq$}.
We study {$\kq$} moments,
rather than {$\fq$} moments,
since the results for different ranks
$q$ are largely independent of each other
as was discussed in section~\ref{sec-moments}.

An early calculation~\cite{bib-webber},
valid to the 
next-to-leading order (n.l.o.) of perturbation theory,
expresses its results in terms of the 
strong coupling strength, $\alpha_S$, 
and the number of active quark flavors in
the parton shower, $n_f$,
allowing the theoretical sensitivity to 
these quantities to be tested.
This calculation does not incorporate energy
conservation into the parton branching processes.
More recently,
a calculation~\cite{bib-analytic1} has been presented
which is exact for a fixed value of~$\alpha_S$
and valid to the next-to-next-to-leading order (n.n.l.o.)
if the coupling strength is allowed to run.
In this paper,
we refer to this result as the ``n.n.l.o.'' calculation.
By ``exact'', it is meant that the QCD evolution equation 
is solved without resorting to a perturbative expansion.
Energy conservation, but not momentum conservation,
is included in the n.n.l.o. result
(angular-ordering of partons,
introduced to partially account for coherence effects,
results in 
approximate momentum conservation~\cite{bib-dreminpriv}).
The n.n.l.o. results are presented for a 
fixed value of the coupling strength,~$\alpha_S$=0.22.
This value is intended to be an ``effective'' one,
appropriate to account for realistic running of $\alpha_S$
in the parton evolution of Z$^0$ decays.
The number of active quark flavors is assumed
to be $n_f$=4.
Therefore, unlike the n.l.o. calculation,
the n.n.l.o. calculation has not yet been presented in
a form which allows the values of $\alpha_S$
and $n_f$ to be varied.
The results of the n.n.l.o. calculation are not believed 
to be strongly dependent on the choice of $\alpha_S$ or~$n_f$
or on the use of a fixed $\alpha_S$ value rather than
a running value, however~\cite{bib-dreminrev,bib-dreminpriv}.

\begin{figure}[ht]
\epsfxsize=15cm
\epsffile[15   100 545 700]{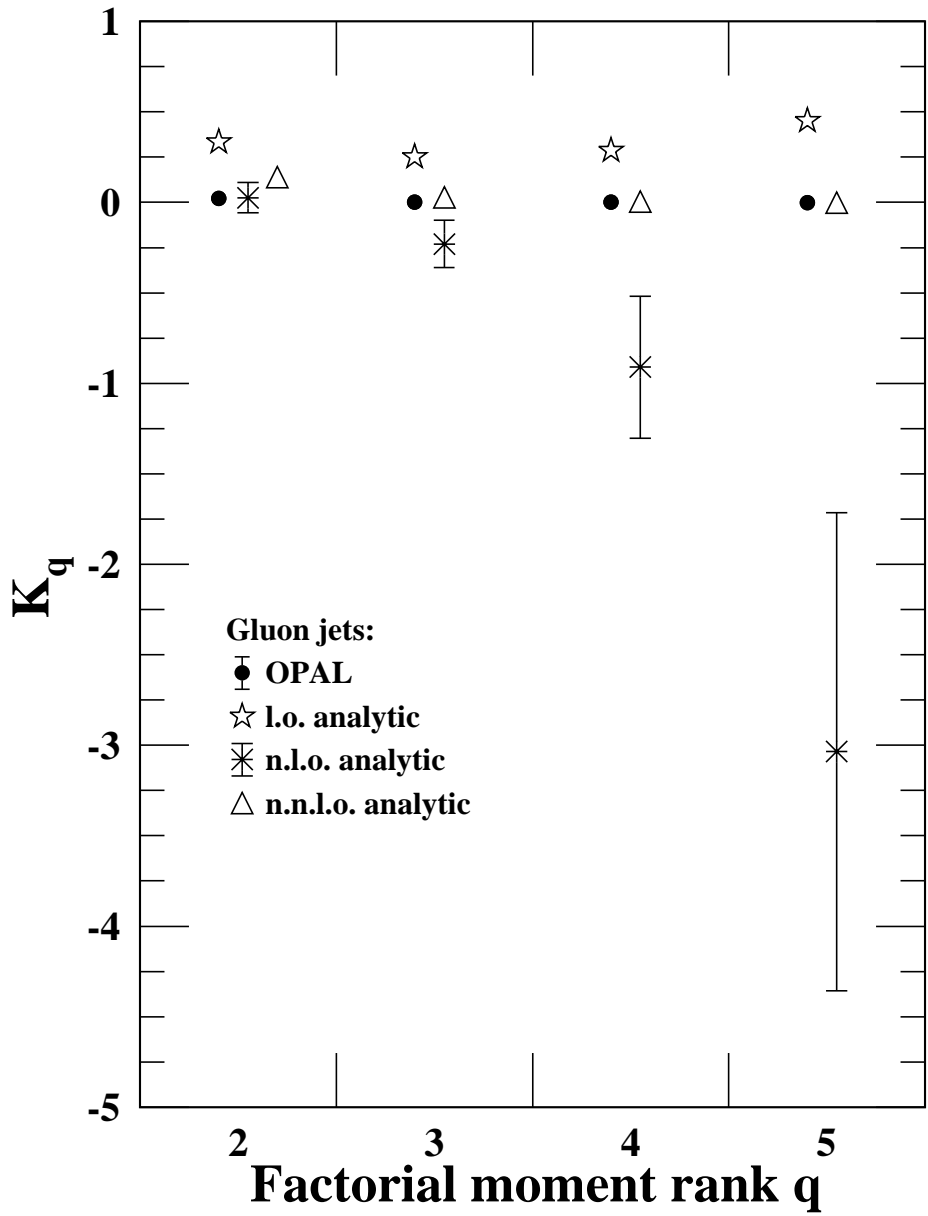}
\caption{
Analytic predictions for the cumulant factorial moments,
{$\kq$}, of gluon jets,
in comparison to the OPAL measurements.
The uncertainties evaluated for the n.l.o.
analytic calculation are described in the text.
}
   \label{fig-ganalytic}
\end{figure}

\begin{figure}[ht]
\epsfxsize=15cm
\epsffile[15   100 545 700]{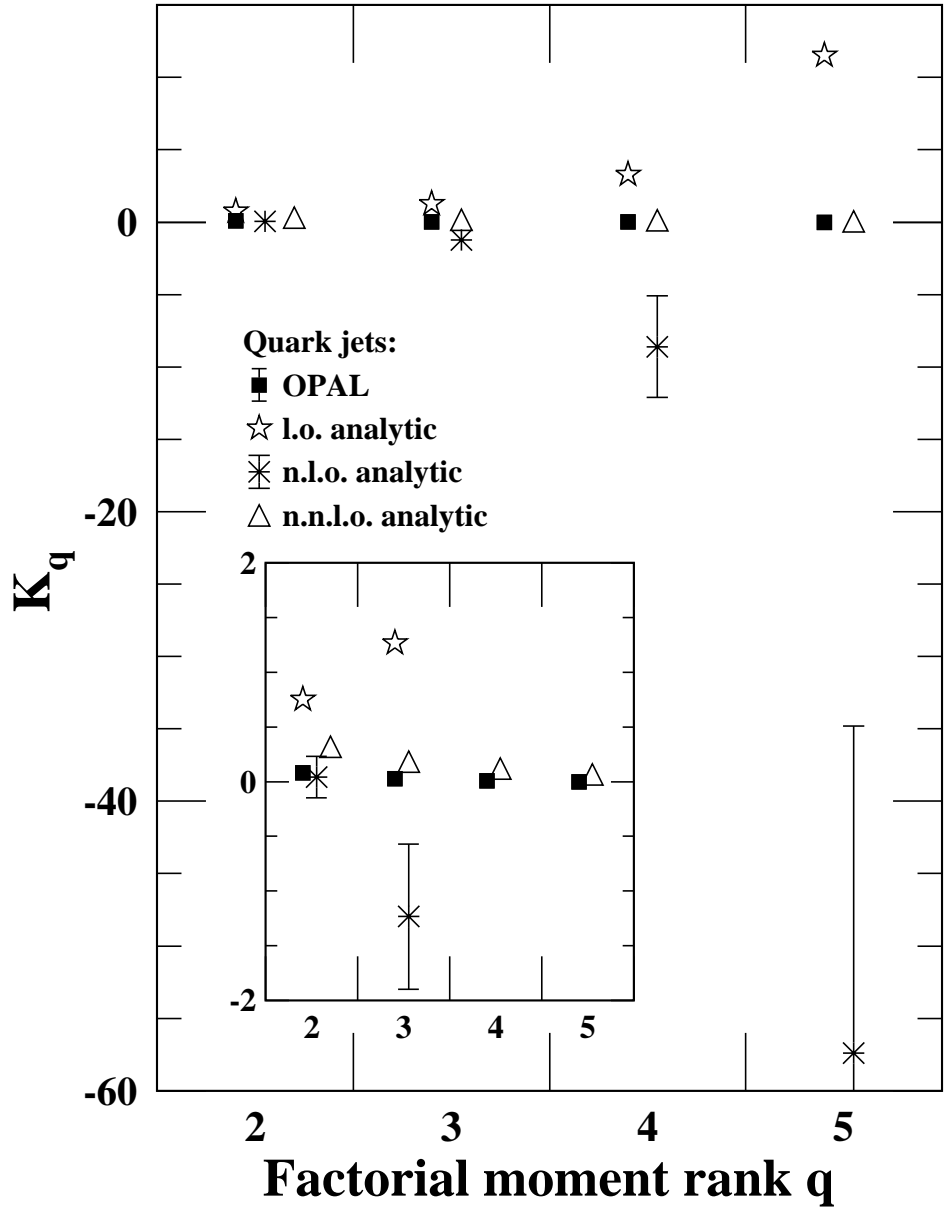}
\caption{
Analytic predictions for the cumulant factorial moments,
{$\kq$}, of quark jets,
in comparison to the OPAL measurements.
The inset shows an expanded view.
The uncertainties evaluated for the n.l.o.
analytic calculation are described in the text.
}
   \label{fig-qanalytic}
\end{figure}

In Figures~\ref{fig-ganalytic} and~\ref{fig-qanalytic},
we present the predictions of the analytic calculations
for the {$\kq$} moments
in comparison to our measurements
from Figure~\ref{fig-kqlow}.
The results are shown for gluon jets 
in Figure~\ref{fig-ganalytic} and for
quark jets in Figure~\ref{fig-qanalytic}.
Besides the n.l.o. and n.n.l.o. results,
we show the leading order (l.o.) results,
obtained from the n.l.o. equations by dropping
the n.l.o. correction terms.
The n.l.o. formulae are evaluated
under three conditions:
(1)~$n_f$=5 and
{\lmsnffive}$\,=0.209$~GeV~\cite{bib-pdg96},
(2)~$n_f$=3 and {\lmsnfthree}$\,=0.340$~GeV,
for which {\lmsnfthree} is
derived from {\lmsnffive} using
the prescription relating {\lmsnfthree}
to {\lmsnffive} given in~\cite{bib-bernreuther},
and (3)~$n_f$=5 and {\lmsnffive}$\,=0.209$~GeV,
with the energy scale at which $\alpha_S$
is evaluated reduced from {\ecm} to 
{\ecm}/4.\footnote{This choice of energy 
scales is taken from~\cite{bib-kn}.}
We take the midpoint between the extreme values
found using these three conditions as the
central n.l.o. result,
and define a theoretical uncertainty by taking
the difference between the central
and extreme values:
the extreme values are in all cases given
by condition~(1),
which yields the maximum predicted 
values of {$\kq$} at n.l.o.,
and~(2), which yields the minimum predicted values.

From Figures~\ref{fig-ganalytic} and~\ref{fig-qanalytic},
it is seen that the predictions of the l.o. calculation
(the star symbols)
are always well in excess of the data.
It is seen that large negative corrections are
introduced at n.l.o.
(the asterisk symbols with uncertainties in
Figures~\ref{fig-ganalytic} and~\ref{fig-qanalytic}),
which bring the theory
into agreement with the data for $q$=2,
but which result in even larger discrepancies
between data and theory for $q$=4 and $q$=5 than
are observed at~l.o.
In contrast,
the n.n.l.o. calculation
(the triangle symbols in
Figures~\ref{fig-ganalytic} and~\ref{fig-qanalytic})
is seen to provide a reasonable qualitative
description of the gluon and quark jet results
for all $q$ values.
There remain important numerical discrepancies between the
n.n.l.o. predictions and our data:
the n.n.l.o. results for {$\kq$} are
0.14, 0.029, 0.0051 and $-$0.00042
for gluon jets and
0.32, 0.18, 0.12 and 0.065
for quark jets~\cite{bib-analytic1},
for ranks $q$=2, 3, 4 and~5, respectively.
For $q$=2 and~3,
these results are typically a factor
of 3 to~6 larger than the experimental results given
in Tables~\ref{tab-gmoments} and~\ref{tab-qmoments};
for $q$=4 and~5,
the discrepancies are in some cases larger 
and in some cases smaller than this.
Nonetheless,
it is apparent from 
Figures~\ref{fig-ganalytic} and~\ref{fig-qanalytic}
that the n.n.l.o. results represent
a striking improvement in the theoretical 
description of the cumulant moment data
in comparison to the results provided by the 
lower order calculations.
This suggests that higher order corrections and energy
conservation are essential to obtain a reasonable
analytic description of
gluon and quark jet multiplicity data,
similar to what we observed in our study
of the mean multiplicity 
ratio~{\rch}~\cite{bib-opalqg96}.

\begin{figure}[ht]
\epsfxsize=15cm
\epsffile[15   100 545 700]{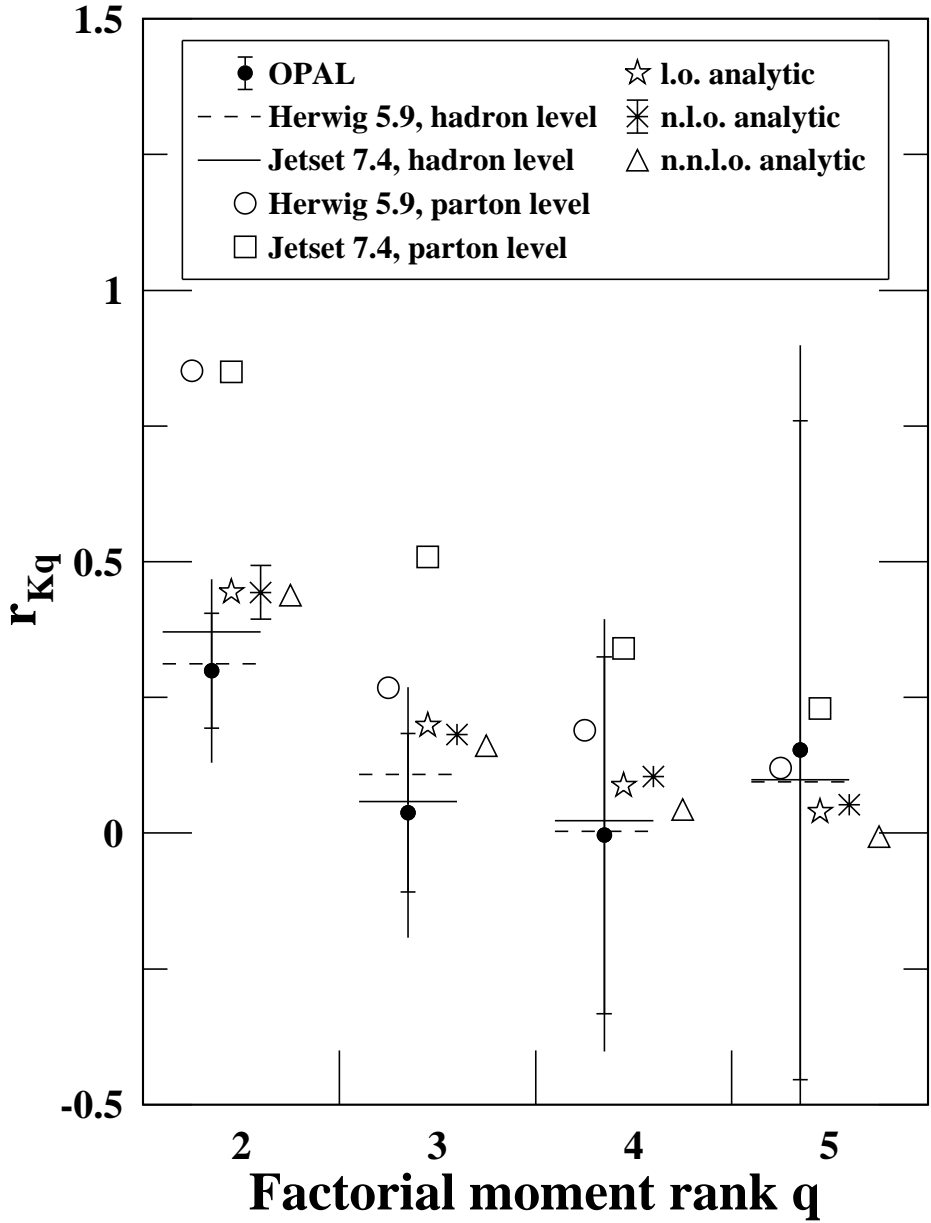}
\caption{
The ratios {$\rkq$} of the cumulant factorial moments 
{$\kq$} of 41.8~GeV gluon and quark jets,
in comparison to the predictions of
QCD analytic calculations and the
Herwig and Jetset
parton shower event generators.
The total uncertainties of the data
are shown by vertical lines.
The experimental statistical uncertainties are
indicated by small horizontal bars.
The uncertainties evaluated for the n.l.o.
analytic calculation are described in the text.
}
   \label{fig-kqratio}
\end{figure}

Although the analytic calculations do not
provide an accurate quantitative description
of the gluon and quark jet moments,
it can be anticipated that certain factors,
such as hadronization and the dependence of 
the predictions on $n_f$ or the energy scale,
will be common to the gluon and quark jet results.
Therefore, in Figure~\ref{fig-kqratio},
we show the analytic predictions 
for the ratios~{$\rkq$},
defined in section~\ref{sec-moments}.
For purposes of comparison,
the parton level predictions of Herwig
and Jetset are shown as well.
It is seen that the three analytic results,
valid to l.o., n.l.o. and n.n.l.o.,
yield almost identical results for~{$\rkq$}.
This agreement,
in stark contrast to the very different predictions 
which the calculations provide for the individual 
gluon and quark jet moments
(Figures~\ref{fig-ganalytic} and~\ref{fig-qanalytic}),
suggests that the theoretical uncertainties
of {$\rkq$} are small,
and in particular that these ratios are only weakly
sensitive to the effect of energy conservation 
and to the values of $\alpha_S$ and $n_f$.
The theoretical uncertainties 
evaluated for the n.l.o. results are seen to be
much larger than the experimental uncertainties in
Figures~\ref{fig-ganalytic} and~\ref{fig-qanalytic},
but much smaller than the experimental uncertainties
in Figure~\ref{fig-kqratio},
which supports this conclusion.

For $q$=2,
the analytic calculations predict that the
{$\kq$} moment of gluon jets is smaller than
the corresponding moment of quark jets
by a factor of about~2.3 (Figure~\ref{fig-kqratio}).
For higher ranks,
the difference between the gluon and quark jet moments
is predicted to be even larger.
These results are in good agreement with
our measurements,
as is seen from Figure~\ref{fig-kqratio}.

Comparing the {$\rkq$} results of Herwig and Jetset
at the parton and hadron levels
(i.e.~comparing the open circle symbols to the
dashed horizontal lines and the square symbols
to the solid horizontal lines in Figure~\ref{fig-kqratio}),
it is seen that the hadronization corrections
predicted by the models are not negligible,
especially for $q$=2 and $q$=3.
The interpretation of the parton level
Monte Carlo predictions is somewhat unclear,
however, for two reasons.
First,
the Monte Carlo simulations implement cutoffs,
denoted {\qzero},
to truncate the parton shower at 
small parton invariant masses.
For Herwig and Jetset, {\qzero}$\,\sim\,$1~GeV.
We find the parton level Monte Carlo results
for~{$\rkq$} to be sensitive to~{\qzero}.
In contrast,
analytic predictions for {$\rkq$}
do not depend on a cutoff
and in this sense are more reliable theoretically.
Second,
the mean numbers of partons present
at the end of the perturbative shower
are small for the Monte Carlo results
shown in Figure~\ref{fig-kqratio}:
an average of only 4.6 and 2.9 partons are present, 
respectively, for 41.8~GeV gluon
and uds quark jet hemispheres generated
using Herwig with our tuned parameter set.
The corresponding results for Jetset
are 4.1 and~2.9.
These small numbers make it questionable whether
the parton level Monte Carlo predictions for {$\rkq$}
have much numerical meaning.
To further investigate this question,
we used Herwig to generate 5~TeV
{\gluglu} and uds {\qq} hemisphere jets:
the resulting mean number of partons present 
at the end of the perturbative shower
was 56.7 for gluon jets 
and 30.2 for quark jets.
The 5~TeV Monte Carlo jets avoid the two problems 
mentioned above for 41.8~GeV Monte Carlo jets:
the condition {\ejet}$\,>>\,${\qzero}
effectively eliminates the dependence of the
parton level predictions on~{\qzero},
while the large parton multiplicities
make calculation of higher factorial 
and cumulant moments numerically sensible.
The ratios {$\rkq$} determined using the
simulated 5~TeV jets were found to be in good 
agreement with the analytic results shown
in Figure~\ref{fig-kqratio}
at both the parton and hadron levels.
We conclude that the hadronization corrections
predicted by the simulations for 41.8~GeV jets are
numerically questionable and that the agreement
between the data and analytic calculations
shown in Figure~\ref{fig-kqratio} is probably
not coincidental.

The OPAL results are based on event hemispheres.
The n.n.l.o. results are based on the 
full event multiplicity distributions
(in contrast, the definitions employed
for the n.l.o. calculation correspond
to hemispheres~\cite{bib-webberpriv}).
To assess the effect of this difference,
we used Herwig and Jetset to evaluate the
{$\kq$} moments of the full event
charged particle multiplicity distributions.
The full event results were compared to the
corresponding hemisphere results shown in
Figures~\ref{fig-kqlow} and~\ref{fig-kqratio}.
Use of full events rather than hemispheres
was found to reduce the magnitude of the
gluon and quark jet {$\kq$} moments,
typically by 30-60\%,
further increasing the quantitative discrepancy
between the data and n.n.l.o. calculation.
This difference does not
affect the qualitative agreement between 
the data and n.n.l.o. calculation exhibited in 
Figures~\ref{fig-ganalytic} and~\ref{fig-qanalytic}
since the visible positions of the data
are already near zero on the scales of those figures.
The effect of the difference between hemispheres
and full events was found to be negligible for the
ratios~{$\rkq$} shown in Figure~~\ref{fig-kqratio}.

\section{Summary and conclusion}
\label{sec-summary}

In this study,
we have presented measurements of the multiplicity
distributions of gluon and quark jets.
We have determined their mean, dispersion, 
skew and curtosis values, 
and factorial and cumulant moments.
The gluon and quark jets are defined by inclusive
sums over the particles in {\gincl} and uds event
hemispheres, respectively,
with the {\gincl} gluon jet opposite to a hemisphere 
containing two identified quark jets in {\epem} annihilations
(the quark jets for the {\gincl} identification
are defined using a jet finding algorithm).
These inclusive definitions are in close
correspondence to the definition of jets used for
QCD analytic calculations,
allowing a meaningful comparison of data with theory.
Our results for the gluon jet
properties are almost independent
of the choice of the jet finding algorithm,
in contrast to other studies of high energy
({\ejet}$>$~5~GeV) gluon jets.
The energy of the jets in our study is about~42~GeV.

We find the mean multiplicity values of gluon and
quark jets to differ by about 50\%,
in agreement with our earlier result~\cite{bib-opalqg96}
but with a substantially reduced uncertainty.
We also observe differences between
the gluon and quark jet skew and curtosis values:
the multiplicity distribution of quark jets is observed
to be about twice as skewed (asymmetric) as
the multiplicity distribution of gluon jets,
while quark jets are found to have a larger curtosis value
(curtosis measures the deviation of a
distribution from a gaussian shape)
than gluon jets.
The dispersions of gluon and quark jets are
found to be the same to within the 
experimental uncertainties.
These results are well reproduced by the predictions
of QCD parton shower event generators.

We analyze the gluon and quark jet distributions
to determine their
normalized factorial and cumulant factorial moments.
These measurements are used to perform the
first test of QCD analytic predictions of
these moments for separated gluon and
quark jets.
We base our test of the analytic results on
the cumulant factorial moments,
{$\kq$}, since we observe that {$\kq$} moments 
of different rank $q$ are largely uncorrelated
with each other,
unlike the factorial moments,~{$\fq$}.
A recent next-to-next-to-leading order calculation 
which includes energy conservation~\cite{bib-analytic1} 
is found to provide a a striking improvement 
in the theoretical description of the individual
gluon and quark jet {$\kq$} values,
in comparison to the description provided by
the leading and next-to-leading order calculations.

Our analysis of the {$\kq$} moments
reveals large differences between
gluon and quark~jets.
For rank $q$=2, the ratio of the cumulant moments 
of gluon to quark jets is found to be
$r_{K_2}$=$0.30\pm0.11\,{\mathrm{(stat.)}}
\pm0.13\,{\mathrm{(syst.)}}$.
For rank $q$=3, 
$r_{K_3}$=$0.04\pm0.15\,{\mathrm{(stat.)}}
\pm0.18\,{\mathrm{(syst.)}}$.
The analytic predictions for $\rkq$ are found
to be in quantitative agreement with the data.

\section{Acknowledgements}

We wish to thank the SL Division 
for the efficient operation
of the LEP accelerator and for
their continuing close cooperation with
our experimental group.  
We thank our colleagues from CEA, DAPNIA/SPP,
CE-Saclay for their efforts over the years on 
the time-of-flight and trigger
systems which we continue to use.  
In addition to the support staff at our own
institutions we are pleased to acknowledge the:

\noindent
Department of Energy, USA, \\
National Science Foundation, USA, \\
Particle Physics and Astronomy Research Council, UK, \\
Natural Sciences and Engineering Research Council, Canada, \\
Israel Science Foundation, administered by the Israel
Academy of Science and Humanities, \\
Minerva Gesellschaft, \\
Benoziyo Center for High Energy Physics,\\
Japanese Ministry of Education, Science and Culture (the
Monbusho) and a grant under the Monbusho International
Science Research Program,\\
German Israeli Bi-national Science Foundation (GIF), \\
Bundesministerium f\"ur Bildung, Wissenschaft,
Forschung und Technologie, Germany, \\
National Research Council of Canada, \\
Hungarian Foundation for Scientific Research, OTKA T-016660, 
T023793 and OTKA F-023259.\\

\pagebreak\clearpage

\pagebreak
\begin{table}[p]
\centering
\begin{tabular}{|l|cc|}
 \hline
  & & \\[-2.4mm]
  {\nch} & P({\nch}), {\gincl} jets (\%) 
         & P({\nch}), uds jets (\%)\\[2mm]
 \hline
 \hline
 0 & ---                  & $0.040\pm0.006\pm0.027$ \\
 1 & ---                  & $0.182\pm0.011\pm0.068$ \\
 2 & ---                  & $0.726\pm0.023\pm0.099$ \\
 3 & ---                  & $1.74\pm0.03\pm0.18$    \\
 4 & $0.31\pm0.20\pm0.38$ & $3.64\pm0.05\pm0.20$    \\
 5 & $0.86\pm0.44\pm0.43$ & $5.59\pm0.05\pm0.25$    \\
 6 & $1.24\pm0.63\pm0.68$ & $8.12\pm0.07\pm0.26$    \\
 7 & $2.08\pm0.85\pm0.77$ & $9.41\pm0.06\pm0.23$    \\
 8 & $3.9\pm1.0\pm1.0$    & $10.54\pm0.07\pm0.26$    \\
 9 & $5.2\pm1.3\pm0.9$    & $10.08\pm0.07\pm0.24$    \\
10 & $6.5\pm1.4\pm0.9$    & $9.57\pm0.07\pm0.27$    \\
11 & $7.0\pm1.3\pm1.1$    & $8.21\pm0.06\pm0.20$    \\
12 & $7.7\pm1.6\pm1.0$    & $7.06\pm0.06\pm0.19$    \\
13 & $9.0\pm1.5\pm1.1$    & $5.69\pm0.05\pm0.12$    \\
14 & $9.4\pm1.5\pm0.8$    & $4.64\pm0.06\pm0.10$    \\
15 & $9.4\pm1.5\pm1.0$    & $3.637\pm0.042\pm0.075$ \\
16 & $8.7\pm1.5\pm1.0$    & $2.858\pm0.036\pm0.081$ \\
17 & $7.2\pm1.4\pm1.1$    & $2.195\pm0.034\pm0.090$ \\
18 & $5.6\pm1.1\pm1.2$    & $1.688\pm0.030\pm0.077$ \\
19 & $4.6\pm1.1\pm1.1$    & $1.254\pm0.027\pm0.062$ \\
20 & $3.2\pm0.9\pm1.0$    & $0.932\pm0.023\pm0.045$ \\
21 & $2.38\pm0.90\pm0.68$ & $0.672\pm0.019\pm0.046$ \\
22 & $1.55\pm0.71\pm0.46$ & $0.477\pm0.016\pm0.062$ \\
23 & $1.08\pm0.81\pm0.48$ & $0.342\pm0.013\pm0.070$ \\
24 & $0.81\pm0.58\pm0.64$ & $0.241\pm0.011\pm0.058$ \\
25 & $1.23\pm0.52\pm0.84$ & $0.167\pm0.010\pm0.034$ \\
26 & $0.27\pm0.30\pm0.86$ & $0.113\pm0.007\pm0.017$ \\
27 & $0.66\pm0.32\pm0.57$ & $0.074\pm0.006\pm0.013$ \\
28 & ---                  & $0.050\pm0.004\pm0.010$\\
29 & $0.05\pm0.14\pm0.18$ & $0.0311\pm0.0035\pm0.0088$\\
30 & $0.15\pm0.09\pm0.15$ & $0.0222\pm0.0029\pm0.0066$\\
31 & ---                  & $0.0130\pm0.0020\pm0.0059$\\
32 & ---                  & $0.0091\pm0.0016\pm0.0039$\\
33 & ---                  & $0.0039\pm0.0012\pm0.0029$\\
34 & ---                  & $0.0024\pm0.0011\pm0.0016$\\
35 & ---                  & $0.0016\pm0.0008\pm0.0012$\\
36 & ---                  & $0.00056\pm0.00039\pm0.00077$\\
 \hline
\end{tabular}
\caption{
Charged particle multiplicity distributions,
expressed in per cent~(\%),
of 41.8~GeV \protect{\gincl} gluon jets
and 45.6~GeV uds quark jets.
The first uncertainty is statistical and the
second is systematic.
The statistical and systematic uncertainties 
are correlated between bins.
}
\label{tab-nchresults}
\end{table}

\pagebreak
\begin{table}[p]
\centering
\begin{tabular}{|c|ccc|}
 \hline
  & & & \\[-2.4mm]
 $q$ & {$\fq$} & {$\kq$} & {$\hq$} \\[2mm]
 \hline
 \hline
2 & $1.023\pm0.008\pm0.011$ & $0.0233\pm0.0083\pm0.0109$    
        & $0.0228\pm0.0078\pm0.0104$    \\
3 & $1.071\pm0.026\pm0.034$ & $0.0010\pm0.0039\pm0.0048$ 
        & $0.0009\pm0.0035\pm0.0045$  \\
4 & $1.146\pm0.059\pm0.074$ & $0.0000\pm0.0023\pm0.0015$ 
        & $0.0000\pm0.0019\pm0.0013$ \\
5 & $1.25\pm0.11\pm0.13$   & $-0.0005\pm0.0018\pm0.0014$ 
        & $-0.0004\pm0.0013\pm0.0011$ \\
 \hline
\end{tabular}
\caption{
The \protect{$\fq$}, \protect{$\kq$} and \protect{$\hq$} 
factorial moments of the charged particle
multiplicity distribution of 41.8~GeV 
\protect{\gincl} gluon jets.
The first uncertainty is statistical and the 
second is systematic.
}
\label{tab-gmoments}
\end{table}

\pagebreak
\begin{table}[p]
\centering
\begin{tabular}{|c|ccc|}
 \hline
  & & & \\[-2.4mm]
 $q$ & {$\fq$} & {$\kq$} & {$\hq$} \\[2mm]
 \hline
 \hline
2 & $1.0820\pm0.0006\pm0.0046$ & $0.0820\pm0.0006\pm0.0048$ 
       & $0.0758\pm0.0005\pm0.0041$ \\
3 & $1.275\pm0.002\pm0.017$  & $0.0291\pm0.0006\pm0.0035$ 
       & $0.0228\pm0.0004\pm0.0026$ \\
4 & $1.637\pm0.005\pm0.042$  & $0.0081\pm0.0007\pm0.0015$  
       & $0.00496\pm0.00043\pm0.00089$  \\
5 & $2.274\pm0.014\pm0.093$  & $-0.00300\pm0.00096\pm0.00095$ 
       & $-0.00132\pm0.00043\pm0.00044$ \\
 \hline
\end{tabular}
\caption{
The \protect{$\fq$}, \protect{$\kq$} and \protect{$\hq$} 
factorial moments of the charged particle
multiplicity distribution of 45.6~GeV uds quark jets.
The first uncertainty is statistical and the 
second is systematic.
}
\label{tab-qmoments}
\end{table}

\pagebreak
\begin{table}[p]
\centering
\begin{tabular}{|c|cccccccccccc|}
 \hline
  & & & & & & & & & & & & \\[-2.4mm]
  & {\mnch} & $D$ &  $\gamma$ & $c$ 
  & $F_2$ & $F_3$ & $F_4$ & $F_5$ 
  & $K_2$ & $K_3$ & $K_4$ & $K_5$ \\[2mm]
 \hline
 \hline
{\mnch} & 1.00 & 0.22 & 0.13 & 0.03 & $-0.03$ 
      & $-0.01$ & 0.01 & 0.01 
      & $-0.03$ &  0.13 & $-0.06$ & $-0.07$ \\
$D$     &      & 1.00 & 0.36 & 0.14 &  \fbox{0.97} &  \fbox{0.96} 
      & \fbox{0.94} & \fbox{0.90} 
      &  \fbox{0.97} &  0.36 & $-0.02$& $-0.14$ \\
$\gamma$&      &      & 1.00 & 0.77 &  0.33 &  0.46 & 0.57 & 0.66
      &  0.33 &  \fbox{0.99} & 0.42 & $-0.18$ \\
$c$     &      &      &      & 1.00 &  0.14 &  0.24 & 0.36 & 0.48
      &  0.14 &  0.78 & \fbox{0.90} & 0.08 \\
$F_2$   &      &      &      &      &  1.00 &  \fbox{0.99} 
      & \fbox{0.96} & \fbox{0.92} 
      & \fbox{1.00} & 0.33 & $-0.01$ & $-0.13$ \\
$F_3$   &      &      &      &      &       &  1.00 & \fbox{0.99} 
      & \fbox{0.96}
      & \fbox{0.99} & 0.46 & 0.06  & $-0.15$ \\
$F_4$   &      &      &      &      &       &       & 1.00 
      & \fbox{0.99}
      & \fbox{0.96} & 0.58 & 0.15 & $-0.15$ \\
$F_5$   &      &      &      &      &       &       &      & 1.00
      & \fbox{0.92} & 0.67 & 0.26 & $-0.13$ \\
$K_2$   &      &      &      & 
      &      &      &      &      &  
         1.00 &  0.33 & $-0.01$& $-0.13$ \\
$K_3$   &      &      &      & 
      &      &      &      &      &       &  1.00 & 0.43 & $-0.19$ \\
$K_4$   &      &      &      & 
      &      &      &      &      &       &       & 1.00 & 0.25 \\
$K_5$   &      &      &      & 
      &      &      &      &      &       &       &      & 1.00 \\
 \hline
\end{tabular}
\caption{
Correlation matrix between the mean {\mnch}, dispersion $D$, 
skew $\gamma$, curtosis $c$,
and the {$\fq$} and {$\kq$} factorial moments of the 
charged particle multiplicity distribution,
for 41.8~GeV {\gincl} gluon jets obtained using 
the Jetset Monte Carlo at the generator level.
Boxes have been drawn around correlation
coefficients which have magnitudes of~0.90 or larger.
}
\label{tab-fqcorr}
\end{table}

\pagebreak\clearpage

\pagebreak
\begin{table}[p]
\centering
\begin{tabular}{|c|ccc|}
 \hline
  & & & \\[-2.4mm]
 $q$ & {$\rfq$} & {$\rkq$} & {$\rhq$} \\[2mm]
 \hline
 \hline
2 & $0.949\pm0.008\pm0.011$  & $0.30\pm0.11\pm0.13$ 
      & $0.32\pm0.11\pm0.14$ \\
3 & $0.850\pm0.021\pm0.026$ & $0.04\pm0.15\pm0.18$ 
      & $0.05\pm0.17\pm0.22$ \\
4 & $0.716\pm0.037\pm0.040$  & $0.00\pm0.33\pm0.22$  
      & $-0.01\pm0.44\pm0.31$ \\
5 & $0.571\pm0.051\pm0.054$  & $0.15\pm0.61\pm0.43$ 
      & $0.27\pm0.96\pm0.72$ \\
 \hline
\end{tabular}
\caption{
The ratios \protect{$\rfq$}, \protect{$\rkq$} and \protect{$\rhq$}
of the \protect{$\fq$}, \protect{$\kq$} and \protect{$\hq$}
factorial moments of 41.8~GeV {\gincl} gluon and uds quark jets.
The first uncertainty is statistical and the second is systematic.
}
\label{tab-rmoments}
\end{table}

\pagebreak
\begin{table}[p]
\centering
\begin{tabular}{|l|ccc|}
 \hline
 & & & \\[-2.4mm]
 & {\mngincl} & {\mnuds}  & {\rch} \\[2mm]
 \hline
 \hline
1. Charged tracks only  
          & $-0.13$ & +0.05   &  $-0.020$ \\
2. Herwig corrections
          & $-0.10$ & +0.01   &  $-0.012$ \\
3. {$p$}$\,>\,$0.20~GeV/$c$  
          & $-0.15$ & $-0.17$ &  +0.010 \\
\hspace*{.5cm} ($\;$$|\cos(\theta_{\mathrm{particle}})|<0.70$
          & $-0.10$ & $-0.04$ & $\;\;-0.004\;$)  \\
4. Cone jet finder   
          & $-0.23$ & ------  & $-0.023$ \\
\hspace*{.5cm} ($\;$JADE-E0 jet finder
          &  +0.17  & ------  & $\;\;$+0.018$\;$) \\
5. $\theta_{\mathrm{jet\,a-thrust}},
    \theta_{\mathrm{jet\,b-thrust}}<65^\circ$
          &  +0.21  & ------  &  +0.022  \\
\hspace*{.5cm} ($\;$$\theta_{\mathrm{jet\,a-jet\,b}}>65^\circ$
          &  +0.19  & ------  &  $\;\;$+0.020$\;$)  \\
6. One track with $b/\sigma_b>2.5$
          &  +0.04  & ------  &  +0.004 \\
7. On-peak data only    
          & $-0.11$ & ------  & $-0.011$ \\
8. 70$^\circ$ cone      
          & ------  & $-0.01$ &  +0.002 \\
9. $b/\sigma_b>2.5$
          & ------  & $-0.02$ &  +0.003 \\
10. Uncertainty of {\egincl}
          & ------  & ------  & $\pm$0.008 \\
\hspace*{.5cm} ($\;$Energy correction factor
          & ------  & ------  & $\;\;\pm$0.004$\;$) \\
 \hline
 \hline
Total systematic uncertainty
          & 0.40   & 0.18    & 0.043   \\
 \hline
\end{tabular}
\caption{
Differences between the results of
the standard analysis and those found by repeating
the analysis with the systematic changes listed,
for the mean charged particle multiplicity value 
{\mnch} of 41.8~GeV {\gincl} gluon jets,
for the {\mnch} value of 45.6~GeV uds quark jets,
and for the ratio {\rch} between the {\mnch} values of
41.8~GeV {\gincl} and uds quark jets.
For item~4, ``jet~a'' and ``jet~b'' refer to the two
tagged quark jets against which the {\gincl} jet
recoils and ``thrust'' refers to the thrust axis.
For items~3, 4, 5 and~10,
the larger of the two listed differences
was assigned as the systematic uncertainty;
the smaller of the two differences is given
in parentheses for purposes of information.
}
\label{tab-rchsyst}
\end{table}

\pagebreak\clearpage

\clearpage

\pagebreak\clearpage

\clearpage

\clearpage

\clearpage

\clearpage

\clearpage

\clearpage

\clearpage

\clearpage

\end{document}